\documentclass[sigconf, screen, noacm]{acmart}
\usepackage{soul}
\usepackage{subfig}
\usepackage{xspace}
\usepackage[dvipsnames]{xcolor}
\usepackage[ruled,lined]{algorithm2e}

\AtBeginDocument{%
  }

\setcopyright{acmlicensed}
\copyrightyear{2026}
\acmYear{2026}
\acmDOI{XXXXXXX.XXXXXXX}

\acmISBN{978-X-XXXX-XXXX-X/XX/XX}

\setcopyright{none}
\renewcommand\footnotetextcopyrightpermission[1]{}
\ifx\figurename\undefined \def\figurename{Figure}\fi
\renewcommand{\figurename}{Fig.}
\newcommand{\para}[1]{\textit{\textbf{#1}} }
\renewcommand{\subparagraph}[1]{\underline{\textit{#1}} }

\newcommand{\Sect}[1]{Sec.~\ref{#1}}
\newcommand{\Fig}[1]{Fig.~\ref{#1}}
\newcommand{\Tbl}[1]{Tbl.~\ref{#1}}

\newcommand{\Eqn}[1]{Eqn.~\ref{#1}}

\newcommand{\specialcell}[2][c]{\begin{tabular}[#1]{@{}c@{}}#2\end{tabular}}

\def\cF{{\mathcal{F}}}

\newcommand{\mode}[1]{\underline{\textsc{#1}}\xspace}

\newcommand{\proj}{\textsc{Deltoris}\xspace}

\definecolor{myorange}{RGB}{255, 212, 121}

\renewcommand{\hl}[1]{{#1}}

\graphicspath{{figures/}}

\begin{document}

\title[\proj: Real-time VLA Inference in Embodied AI]{\proj: Enabling Real-time VLA Inference in Embodied AI via Bit-level Sparsity and Speculative Inference}
% \subtitle{\normalsize{MICRO 2026 Submission
%     \textbf{\#796} -- Confidential Draft -- Do NOT Distribute!!}}
%\author{\normalsize{MICRO 2026 Submission
 %   \textbf{\#NaN} -- Confidential Draft -- Do NOT Distribute!!}}

\author{Zheng Liu}
\authornote{Equal contribution.}
\orcid{0009-0001-6688-4115}
\affiliation{%
  \institution{Shanghai Jiao Tong University}
  \city{Shanghai}
  \country{China}
}
\email{distilledw@sjtu.edu.cn}

\author{Zeyu Guo}
\authornotemark[1]
\orcid{}
\affiliation{%
  \institution{Shanghai Jiao Tong University}
  \city{Shanghai}
  \country{China}
}
\email{zeyuguo77@sjtu.edu.cn}

\author{Zihan Liu}
\orcid{0000-0002-0874-0682}
\affiliation{%
  \institution{Shanghai Jiao Tong University, Shanghai Qi Zhi Institute}
  \city{Shanghai}
  \country{China}
}
\email{altair.liu@sjtu.edu.cn}

\author{Anbang Wu}
\orcid{}
\affiliation{%
  \institution{Shanghai Jiao Tong University}
  \city{Shanghai}
  \country{China}
}
\email{anbang@cs.sjtu.edu.cn}

\author{Han Zhao}
\authornote{Corresponding Authors.}
\orcid{????}
\affiliation{%
  \institution{Shanghai Jiao Tong University}
  \city{Shanghai}
  \country{China}
}
\email{zhao-han@cs.sjtu.edu.cn}

\author{Fangxin Liu}
\orcid{????}
\affiliation{%
  \institution{Shanghai Jiao Tong University}
  \city{Shanghai}
  \country{China}
}
\email{liufangxin@sjtu.edu.cn}

\author{Zhezhi He}
\orcid{0000-0002-6357-236X}
\affiliation{%
  \institution{Shanghai Jiao Tong University}
  \city{Shanghai}
  \country{China}
}
\email{zhezhi.he@sjtu.edu.cn}

\author{Yinhe Han}
\orcid{???}
\affiliation{%
  \institution{ICT, Chinese Academy of Sciences}
  \city{Beijing}
  \country{China}
}
\email{yinhes@ict.ac.cn} 

\author{Jingwen Leng}
\orcid{0000-0002-5660-5493}
\affiliation{%
  \institution{Shanghai Jiao Tong University, Shanghai Qi Zhi Institute}
  \city{Shanghai}
  \country{China}
}
\email{leng-jw@sjtu.edu.cn}

\author{Minyi Guo}
\orcid{0000-0003-0034-2302}
\affiliation{%
  \institution{Shanghai Jiao Tong University, Shanghai Qi Zhi Institute}
  \city{Shanghai}
  \country{China}
}
\email{guo-my@cs.sjtu.edu.cn}

\author{Yiming Gan}
\authornotemark[2]
\orcid{0000-0002-2033-5057}
\affiliation{%
  \institution{ICT, Chinese Academy of Sciences}
  \city{Beijing}
  \country{China}
}
\email{ganyiming@ict.ac.cn}

\author{Yu Feng}
\authornotemark[2]
\orcid{0000-0002-2192-5737}
\affiliation{%
  \institution{Shanghai Jiao Tong University, Shanghai Qi Zhi Institute}
  \city{Shanghai}
  \country{China}
}
\email{y-feng@sjtu.edu.cn}

\renewcommand{\shortauthors}{Zheng Liu et al.} 


%%%%%% -- PAPER CONTENT STARTS-- %%%%%%%%

\begin{abstract}

Vision-language-action (VLA) models have emerged as a key component in embodied AI. 
Among existing approaches, diffusion-based VLA models achieve superior motion quality and generalization.
However, diffusion-based VLA models are compute-intensive and must run at high control frequency, e.g., 50–200 Hz.
Thus, it imposes strict latency and energy constraints on edge devices. 

In this work, we present \textsc{Deltoris}~\footnote{The name \textsc{Deltoris} combines \textit{Delta}, reflecting our differential computation, with \textit{Dolores}, the iconic character from \textit{Westworld}.}, an algorithm–hardware co-design framework for efficient diffusion-based VLA inference. 
First, we exploit the temporal similarity of consecutive inputs and propose a \textit{temporal-aware bit-sparsity} algorithm that computes only the differences between consecutive inputs, eliminating redundant bit-level operations. 
To further address the extra off-chip traffic introduced by our algorithm, we propose a \textit{speculative inference} technique, which amortizes data loading across multiple control steps. 
Lastly, to support these techniques, we co-design a dedicated accelerator with customized 1D systolic bit-serial PE arrays that eliminate PE workload imbalance. 
Our evaluation shows that \textsc{Deltoris} achieves up to \hl{34.2$\times$} speedup over mobile GPUs and \hl{6.1$\times$} over prior accelerators, while maintaining comparable accuracy.

\end{abstract}

%\begin{CCSXML}
%<ccs2012>
%  <concept_id>00000000.0000000.0000000</concept_id>
%  <concept_desc>Do Not Use This Code, Generate the Correct Terms for Your Paper</concept_desc>
%  <concept_significance>500</concept_significance>
%  %<concept_id>00000000.00000000.00000000</concept_id>
%  <concept_desc>Do Not Use This Code, Generate the Correct Terms for Your Paper</concept_desc>
%  <concept_significance>300</concept_significance>
%  %<concept_id>00000000.00000000.00000000</concept_id>
%  <concept_desc>Do Not Use This Code, Generate the Correct Terms for Your Paper</concept_desc>
%  <concept_significance>100</concept_significance>
 % <concept_id>00000000.00000000.00000000</concept_id>
%  <concept_desc>Do Not Use This Code, Generate the Correct Terms for Your Paper</concept_desc>
%  <concept_significance>100</concept_significance>
%</ccs2012>
%\end{CCSXML}

%\ccsdesc[500]{Do Not Use This Code~Generate the Correct Terms for Your Paper}
%\ccsdesc[300]{Do Not Use This Code~Generate the Correct Terms for Your Paper}
%\ccsdesc{Do Not Use This Code~Generate the Correct Terms for Your Paper}
%\ccsdesc[100]{Do Not Use This Code~Generate the Correct Terms for Your Paper}


\maketitle

\section{Introduction}
\label{sec:intro}

Recent advances in embodied artificial intelligence (AI) have enabled robots to perform complex tasks in real-world environments~\cite{robot2026, china_robot2025, boston2026, unitree2026, tesla2026}.
At the core of these systems are vision-language-action (VLA) models, which integrate visual perception, instruction understanding, and action generation into a unified framework.
By leveraging multimodal inputs, VLA models allow robots to interpret high-level instructions and convert them into control commands.

Overall, the architecture of embodied agents resembles the organization of the biological brain.
Large language models (LLMs) function as the cerebrum, responsible for high-level reasoning and planning, while VLA models act as the cerebellum, performing motion planning and generating low-level controls that drive robot behavior.
Thus, these two have different system requirements.
LLM inference is typically event-driven, invoked only when the system needs to perform high-level reasoning, and often operates at low frequencies, e.g., 1–5 Hz~\cite{liu2024okrobot, zhi2025closed, torne2026mem}.
In contrast, VLA models operate within a closed-loop control system, continuously generating actions at much higher frequencies, e.g., 50–200 Hz, to maintain responsive robot motion~\cite{guo2024prediction, chi2023diffusion, li2025unified, hu2024video, belkhale20rth, liu2024rdt, li2023vision, kim24openvla}.

Recent architectural research efforts have primarily focused on accelerating LLM inference via quantization~\cite{xiao2023smoothquant, lin2023awq, frantar2022gptq}, KV-cache compression~\cite{hooper2024kvquant, dong2024qaq, chang2024palu}, sparse attention~\cite{kwon2023efficient, dao2022flashattention, xi2025sparse}, etc.
In contrast, VLA models remain largely \textit{under-explored} in the architecture community.
Currently, enabling VLA models on robotic platforms in real-time is challenging due to their latency and energy constraints.
Existing robotic platforms often fail to meet the real-time requirements (50–200 Hz).
For instance, a well-known VLA model, PAD~\cite{guo2024prediction}, can only execute at 2.0~Hz on a mobile Nvidia Orin SoC (\Fig{fig:vla_latency}).
Thus, this paper aims to enable real-time and energy-efficient VLA model inference on robotic platforms.

Among the VLA models~\cite{guo2024prediction, chi2023diffusion, li2025unified, hu2024video, du2023learning, belkhale20rth, liu2024rdt, li2023vision, zhang2025up, kim24openvla, black2026pi, xu2024flow, wen2024diffusionvla, wen2025dexvla} proposed in recent years, diffusion-based models~\cite{guo2024prediction, chi2023diffusion, li2025unified, hu2024video, du2023learning, belkhale20rth, liu2024rdt, wen2024diffusionvla, wen2025dexvla} have emerged as a promising approach for robot control.
Compared to other inference paradigms, diffusion-based VLA models can produce smooth motion trajectories and achieve better generality among unseen environments by predicting the action distribution through an iterative noise-denoising process.
Recent works, such as Diffusion Policy~\cite{chi2023diffusion}, UVA~\cite{hu2024video}, VPP~\cite{hu2024video}, and Diffusion-VLA~\cite{wen2024diffusionvla} demonstrate that diffusion-based approaches can outperform traditional auto-regressive (AR) policies in many robotic manipulation benchmarks.
In this paper, we primarily focus on diffusion-based VLA models.

\para{Motivation.}
Despite their advantages, diffusion-based VLA models are compute-intensive.
They require tens of iterative denoising steps to generate one individual action, resulting in a higher latency compared to their AR counterparts.
Each denoising step involves executing a deep neural network (DNN) to progressively refine a noisy trajectory into a valid control signal.
Our roofline analysis in  \Fig{fig:roofline} shows that existing diffusion-based VLA models are compute-bound rather than memory-bound.
Thus, it is important to reduce the overall computation of diffusion-based VLA models.

To do so, our key insight is that consecutive robot observations, e.g., poses and images, in robotic control loops exhibit extremely high similarity, given the fact that robot control operates at a high frequency.
Our experiment in \Fig{fig:temporal_similarity} shows that over 98\% of pixels between consecutive camera frames often remain unchanged due to the small time interval between controls.
As a result, a large portion of the computation during VLA inference is redundant.
Potentially, we can enable orders-of-magnitude speedup for diffusion-based VLA inference by exploiting this temporal redundancy.

\para{Framework.}
To exploit this insight, we propose \proj, an algorithm–hardware co-design framework for accelerating diffusion-based VLA models.
\proj exploits the high temporal similarity inherent in robotic control loops and introduces a \textit{temporal-aware bit-level sparsity} algorithm that processes only the differences between consecutive inputs (\Sect{sec:sparsity}).
Similar to the execution of spiking neural networks~\cite{yamazaki2022spiking}, \proj represents differential inputs in binary form and executes computations only on active ``1'' bits while skipping redundant operations.
Our experiment shows in \Fig{fig:perf} that we can achieve up to 92.9\% operation reduction.

Although our bit sparsity algorithm significantly reduces arithmetic operations, it introduces additional off-chip data traffic due to the extra intermediate data loading.
We show that a naive bit sparsity implementation can increase the off-chip data traffic by 1.8$\times$.
We found that the VLA workloads essentially shift from compute-bound to memory-bound after applying our bit-level sparsity algorithm.
To mitigate the DRAM traffic overhead, we propose a technique called \textit{speculative inference} in \Sect{sec:spec}, inspired by speculative decoding in LLMs~\cite{liu2024online, leviathan2023fast}.
In our approach, a lightweight VLA model first speculatively predicts multiple future actions, while a more accurate but compute-intensive VLA model validates these predictions. 
To maintain robustness, no actions are actually applied to the robots before being validated by the larger VLA model.
In this way, both model weights and intermediate data of the larger model can be loaded only once when validating all predicted poses.
Thus, our approach amortizes data loading across multiple validation steps, reducing off-chip data traffic.

\para{Architecture.}
To fully unlock the potential of \proj, we design a dedicated accelerator tailored for diffusion-based VLA inference in \Sect{sec:arch}.
The key contribution of our design is to eliminate the workload imbalance commonly observed in bit-serial accelerators that exploit bit-level sparsity.
Specifically, we employ an output-stationary dataflow that shares activations across multiple processing elements (PEs). 
We further introduce a novel 1D systolic bit-serial PE array.
Compared with prior bit-serial accelerators, our proposed PE array largely simplifies control logic and improves both performance and energy efficiency.
Combined with our co-designed encoding and decoding modules, our architecture effectively reduces both on-chip and off-chip data movement.

\para{Results.} 
We implement an algorithm–hardware co-designed prototype of \proj and evaluate it on three widely used VLA models~\cite{guo2024prediction, chi2023diffusion, li2025unified}. 
Experimental results show that \proj maintains a comparable task success rate with minimal accuracy degradation. 
To demonstrate the performance benefits of \proj, we evaluate our approach on both bit-serial accelerators~\cite{albericio2017bit, chen2024bbs} and diffusion accelerators~\cite{heo2025exion, kong2024cambricon, kim2025ditto}.
Our results show that \proj achieves up to \hl{34.2$\times$ and 6.1$\times$} speedup compared with off-the-shelf GPUs and dedicated accelerators, respectively.

The contributions of this paper are summarized as follows:
\begin{itemize}
\item To the best of our knowledge, \proj is the first algorithm-hardware co-design targeting diffusion-based VLA models.
\item We propose \proj, which exploits the inherent temporal redundancy in robotic control and accelerates VLA inference via bit-level sparsity and speculative inference.
\item We co-design a bit-serial architecture with a tailored dataflow that eliminates the workload imbalance that commonly exists in bit-serial accelerators.
\item \proj achieves up to \hl{34.2$\times$ speedup and 822.0$\times$ energy savings} compared with GPUs and prior accelerators.
\end{itemize}

\section{Background}
\label{sec:bg}

In this section, we first introduce the key concepts of VLA models in \Sect{sec:bg:vla}. We then explain the diffusion process in diffusion-based VLA models in \Sect{sec:bg:diffusion}.

\begin{figure}
    \centering
    \includegraphics[width=\columnwidth]{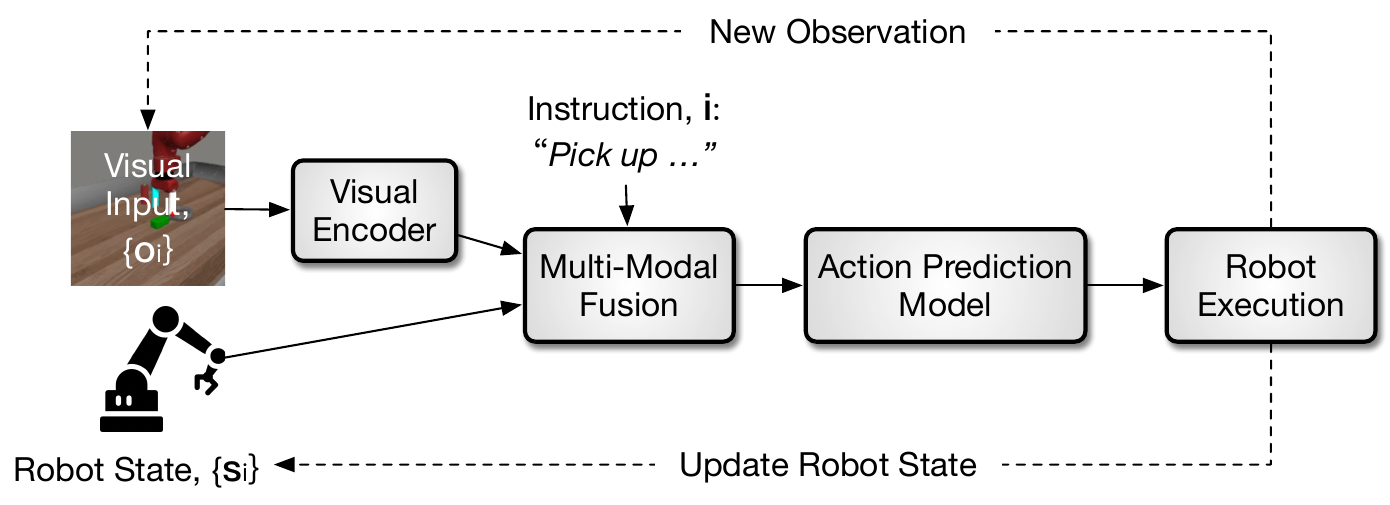}
    \caption{The inference workflow of a VLA model. The vision encoder extracts visual features, and the multi-modal module fuses visual and language tokens to better understand the task context.
    Finally, the action generation module predicts robot control signals that trigger robot execution.}
    \label{fig:vla_workflow}
\end{figure}

\subsection{VLA Model}
\label{sec:bg:vla}

VLA models have recently emerged as a promising paradigm for embodied AI.
By jointly modeling visual perception, instruction understanding, and action generation, VLA models enable robots to translate high-level instructions into executable control commands.

\para{Workflow.}
\Fig{fig:vla_workflow} illustrates the typical inference workflow of a VLA model.
The inputs to a VLA model include camera observations $\{\mathbf{o}_i\}$, natural language instructions $\mathbf{i}$, and robot states $\{\mathbf{s}_i\}$.
Depending on the VLA architecture, the camera observations and robot states may correspond either to the current observation and robot state or to a short history of observations and states, which provides more temporal context.

To consume these inputs, a typical VLA model consists of three main components: a vision encoder, a multi-modal fusion module, and an action generation module.
The vision encoder extracts visual features from camera observations, $\{ \textbf{o}_i \}$, and encodes them in the latent space.
The multi-modal module then fuses visual tokens and language tokens to combine the contexts from different modalities.
Next, the action generation module takes the fused tokens and predicts robot control signals, e.g., joint angles, gripper states, etc.

During robot execution, VLA models operate within a closed-loop control pipeline (\Fig{fig:vla_workflow}).
At each control step, the robot first observes the environment, performs VLA inference to predict an action, and executes the predicted action before receiving the next observation.
This process repeats until the task is complete. 
To maintain safe and smooth behavior of the robot, the VLA inference should run at a high frequency of no less than 50 Hz. 

%Typically, VLA inference runs at high control frequencies, ranging from 50–200 Hz, to maintain safe and smooth behavior.

\begin{figure}[t]
\centering
\begin{minipage}[t]{0.48\columnwidth}
  \centering
  \includegraphics[width=\columnwidth]{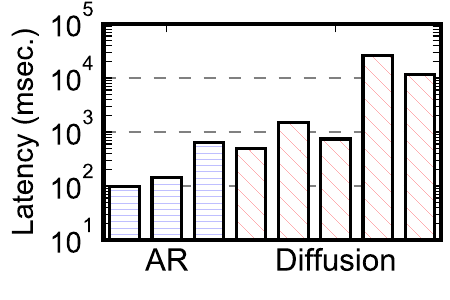}
  \caption{End-to-end latency of various VLA models. We compare diffusion-based VLA~\cite{guo2024prediction, chi2023diffusion, li2025unified, hu2024video, du2023learning} and autoregressive (AR)-based~\cite{wen2025tinyvla, kim2025fine, wang2025unified}.}
  \label{fig:vla_latency}
\end{minipage}
\hspace{2pt}
\begin{minipage}[t]{0.48\columnwidth}
  \centering
  \includegraphics[width=\columnwidth]{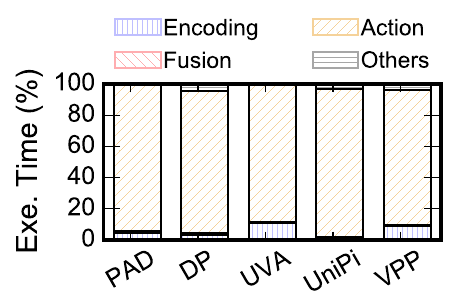}
  \caption{Execution breakdown of diffusion-based VLA models~\cite{guo2024prediction, chi2023diffusion, li2025unified, hu2024video, du2023learning}. Diffusion action prediction dominates the execution.}
  \label{fig:exec_breakdown}
\end{minipage}
\end{figure}

\para{Diffusion-based VLA.}
Among the various VLA architectures proposed in recent years~\cite{guo2024prediction, chi2023diffusion, li2025unified, hu2024video, du2023learning, belkhale20rth, liu2024rdt, li2023vision, zhang2025up, kim24openvla, black2026pi, xu2024flow, wen2024diffusionvla, wen2025dexvla}, diffusion-based VLA models have gained increasing attention due to two key advantages.
First, diffusion-based models can produce smooth and stable action trajectories~\cite{chi2023diffusion} by iteratively refining noisy action sequences.
Thus, it can achieve more physically consistent motions than other VLA models.
Second, by modeling the distribution of robot actions via an iterative diffusion process, these models can capture multi-modal action spaces, enabling better generality across unseen environments~\cite{liu2024rdt, wen2025dexvla}.

However, these performance benefits come at the cost of higher computational overhead.
Unlike AR-based VLA models, which generate actions in a single forward inference, diffusion-based VLA models require multiple iterative denoising steps, with each step invoking the same neural network to refine the predicted action.
This iterative process largely increases inference latency.
As shown in \Fig{fig:vla_latency}, we compare the end-to-end latency of representative VLA models across different inference paradigms, including AR-based and diffusion-based models.
The latency numbers are measured on a mobile Nvidia Orin SoC~\cite{orinsoc}.
Between these two types of models, diffusion-based VLA models show much longer execution time than AR-based models.
In some cases, diffusion models have an order-of-magnitude higher latency than AR-based models.

Moreover, \Fig{fig:exec_breakdown} shows the execution breakdown of diffusion-based VLA models.
The results show that the action prediction, i.e., the diffusion process, dominates the overall inference time.
Across five diffusion-based models, 91.1\% of the total execution time is spent in the diffusion process.
Together, these two results show that accelerating the diffusion process is key to addressing the performance bottleneck in diffusion-based VLA models.

\subsection{Diffusion Process}
\label{sec:bg:diffusion}

\begin{figure}
    \centering
    \includegraphics[width=\columnwidth]{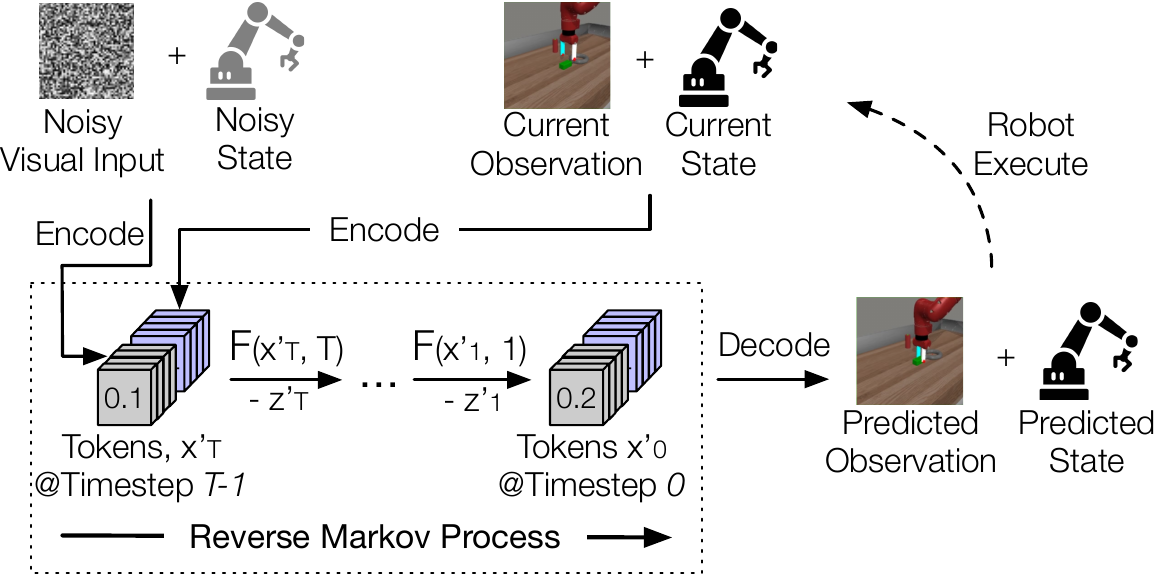}
    \caption{The diffusion process in a diffusion-based VLA model. The noisy inputs and current observations are first encoded into the latent space. Then, the tokens go through $T$ steps of a noising removal process to generate the predicted observation and state, which are then used for action generation.}
    \label{fig:denoising_process}
\end{figure}

Diffusion models~\cite{croitoru2023diffusion, yang2023diffusion, ho2020denoising, kong2024hunyuanvideo, wan2025} are a class of generative models that generate data samples through an iterative noise removal process.
During training, a forward diffusion process gradually perturbs the original data by adding Gaussian noise over multiple steps.
The diffusion model then learns the corresponding reverse diffusion process, which progressively removes noise from corrupted samples to recover the original data distribution.

During VLA inference, diffusion models generate outputs through a sequence of reverse diffusion steps, as shown in \Fig{fig:denoising_process}.
Starting from an initial noisy sample $x’_T$, the model repeatedly applies a denoising network to progressively refine the sample toward the target distribution.
Here, $T$ denotes the total number of denoising steps.
At each step $t$, a diffusion model predicts the noise component, $z’_t$, in the current sample, $x’_t$, and the predicted noise $z’_t$ is then removed from $x’_t$ to obtain a cleaner sample, $x’_{t-1}$.
After $T$ iterations of this denoising process, the model produces the final output $x’_0$, which approximates the original data sample $x_0$.
The formal expression of each denoising step can be described as,
\begin{equation}
x’_{t-1} = \alpha_t (x’_t - \beta_t z’_t) + \sigma_t n’_t,
\quad z’_t = \cF(x’_t, t),
\end{equation}
where $\cF(\cdot)$ denotes the diffusion model that predicts the noise component $z’_t$.
$\alpha_t$ and $\beta_t$ are predefined hyper-parameters that control the denoising process, while $\sigma_t n’_t$ represents a stochastic re-noising term that maintains sample diversity.
After $T$ denoising steps, the final denoised result $x’_0$ would be the output.

In diffusion-based VLA models, the diffusion process is used to generate robot action trajectory or state, $\{\mathbf{s}_i\}$.
Given the current observation $\{\mathbf{o}_i\}$ and task instruction $\mathbf{i}$, the diffusion model predicts a sequence of robot actions $\{\mathbf{s}_i\}$ by progressively refining a noisy action trajectory into a valid control signal.
The final output typically corresponds to low-level robot commands.

\section{Motivation}
\label{sec:motiv}

% \begin{figure}[t]
% \centering
% \begin{minipage}[t]{0.48\columnwidth}
%   \centering
%   \includegraphics[width=\columnwidth]{roofline}
%   \caption{Roofline analysis of diffusion-based VLA models~\cite{guo2024prediction, chi2023diffusion, li2025unified, hu2024video, du2023learning} on Orin.}
%   \label{fig:roofline}
% \end{minipage}
% \hspace{2pt}
% \begin{minipage}[t]{0.48\columnwidth}
%   \centering
%   \includegraphics[width=\columnwidth]{temporal_similarity}
%   \caption{Temporal similarity of adjacent frames in the VLA dataset, PushT~\cite{gym-pusht}.}
%   \label{fig:temporal_similarity}
% \end{minipage}
% \end{figure}

\begin{figure}[t]
\centering
\subfloat[\hl{Orin SoC.}]{
    \label{fig:roofline_orin}
    \includegraphics[width=0.48\columnwidth]{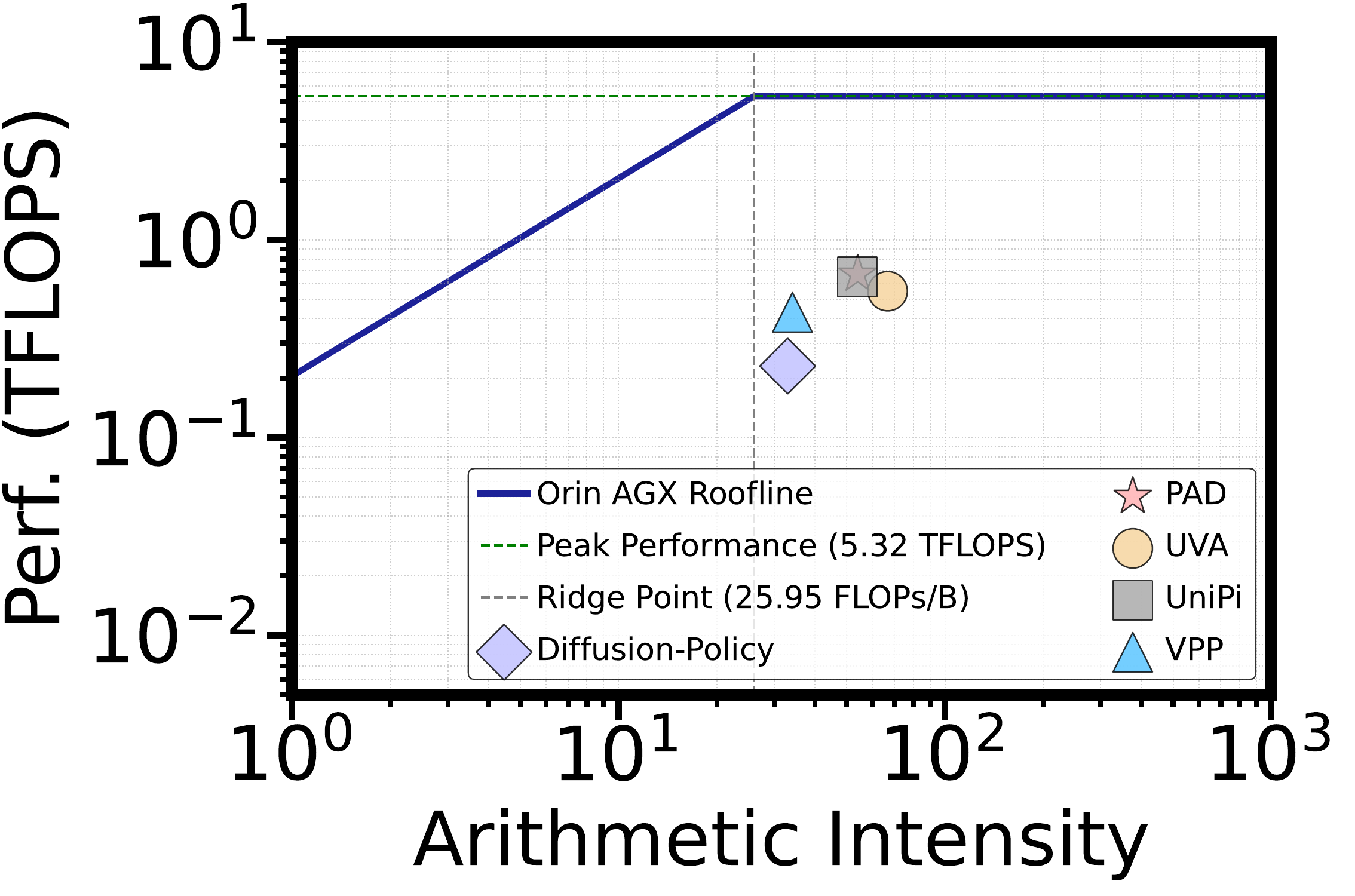}}
\subfloat[\hl{Thor SoC.}]{
    \label{fig:roofline_thor}
    \includegraphics[width=0.48\columnwidth]{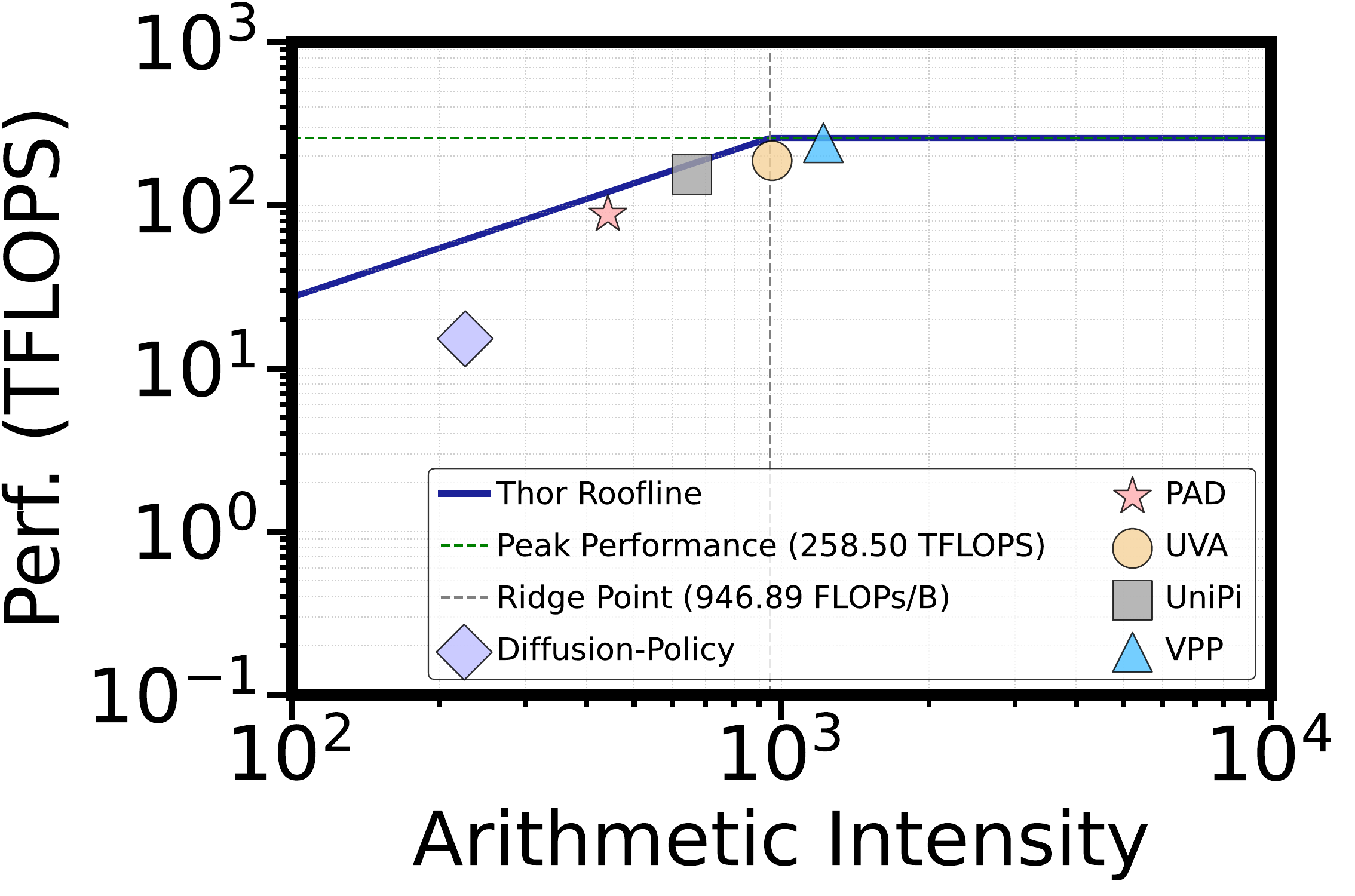}}
\caption{\hl{Roofline analysis of diffusion-based VLA models~\mbox{\cite{guo2024prediction, chi2023diffusion, li2025unified, hu2024video, du2023learning}} on two recent mobile SoCs.}}
\label{fig:roofline}
\end{figure}

\begin{figure*}
    \centering
    \includegraphics[width=\textwidth]{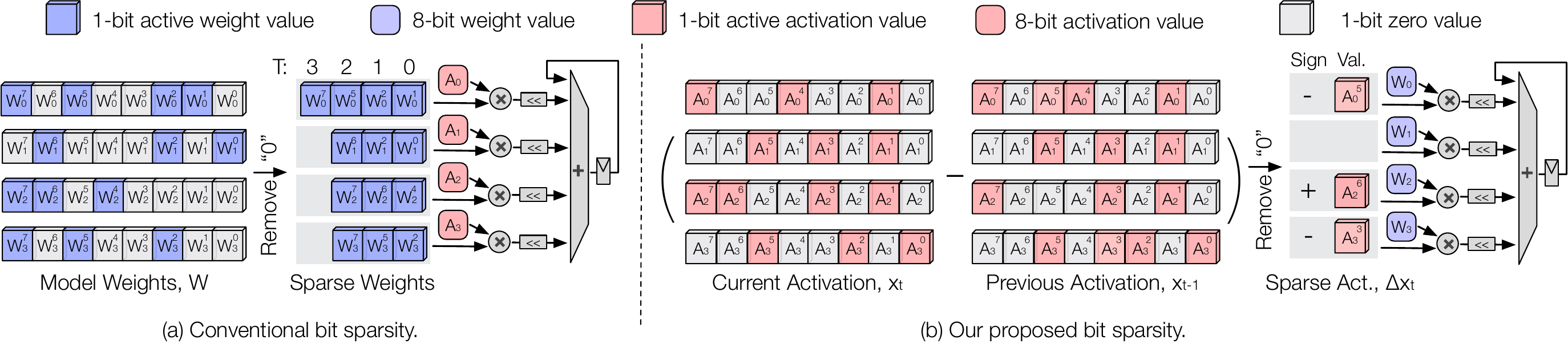}
    \caption{Comparison between conventional bit sparsity and our proposed temporal-aware bit sparsity. Left: conventional bit-sparse execution exploits sparsity only in model weights $\mathbf{W}$ by removing zero-valued bits before computation. Right: our method leverages temporal redundancy between consecutive activations, $\mathbf{x}_t$ and $\mathbf{x}_{t-1}$, and computes only the differential activation $\Delta \mathbf{x}_t = \mathbf{x}_t - \mathbf{x}_{t-1}$. Since $\Delta \mathbf{x}_t$ is much sparser than $\mathbf{W}$ or $\mathbf{x}_{t}$, our approach can reduce more bit operations.}
    \label{fig:bit_sparsity}
\end{figure*}

In \Sect{sec:bg:vla}, we have shown that it is challenging to deploy diffusion-based VLA models on real-world robotic platforms in real-time despite their strong performance benefits.
In this section, we characterize representative diffusion-based VLA models and make two key observations that guide our hardware-algorithm co-design.

\hl{\textit{Observation 1: Diffusion-based VLA inference is primarily compute-bound.}
In \mbox{\Sect{sec:bg:diffusion}}, we show that diffusion-based VLA inference performs a sequence of iterative denoising steps, where each step invokes a diffusion model to progressively refine the action trajectory.
This repeated process introduces substantial arithmetic cost and makes the entire VLA inference compute intensive.

In \mbox{\Fig{fig:roofline}}, we show a roofline analysis of diffusion-based VLA models on two mobile SoCs: Nvidia Orin SoC~\mbox{\cite{orinsoc}} and Nvidia Thor SoC~\mbox{\cite{thorsoc}}.
\mbox{\Fig{fig:roofline}} plots the computation intensity of five diffusion-based VLA models~\mbox{\cite{guo2024prediction, chi2023diffusion, li2025unified, hu2024video, du2023learning}}.
\mbox{\Fig{fig:roofline_orin}} shows that, on Orin SoC, all models are in the compute-bound region of the roofline.
This indicates that their performance is primarily constrained by arithmetic throughput rather than memory bandwidth.
Even on a more powerful Thor SoC, \mbox{\Fig{fig:roofline_thor}} shows that these models are around the boundary between the compute-bound and memory-bound regions.
Thus, for diffusion-based VLAs, reducing the arithmetic workload is an important direction to boost performance.
}

\begin{figure}
    \centering
    \includegraphics[width=\columnwidth]{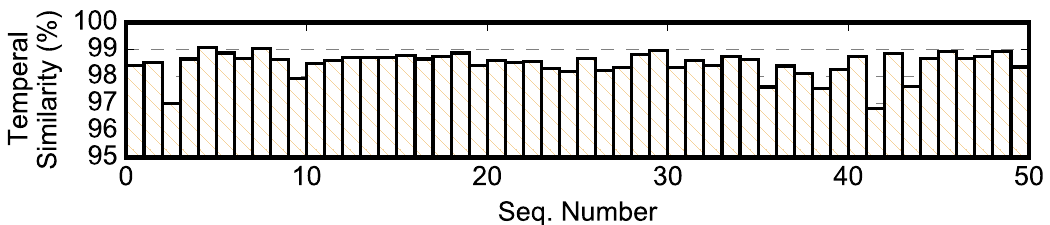}
    \caption{Temporal similarity of adjacent frames in the VLA dataset, PushT~\cite{gym-pusht}.}
    \label{fig:temporal_similarity}
\end{figure}

\textit{Observation 2: Consecutive control steps exhibit strong temporal similarity.}
Although diffusion-based VLA inference is compute-intensive, we find that robotic control loops also exhibit a unique property: the inputs to adjacent control steps are highly similar.
In closed-loop robot execution, observations, robot states, and actions often change gradually over time.
Because the control frequency typically ranges from 50–200 Hz, the time interval between two consecutive inference steps is very small, and the corresponding sensory inputs and robot states often change only slightly.

In \Fig{fig:temporal_similarity}, we quantify the temporal similarity between adjacent images in a widely used robotic dataset, PushT~\cite{gym-pusht}.
Specifically, we compute the ratio of pixel-wise difference between two consecutive frames.
As shown in \Fig{fig:temporal_similarity}, over 98.5\% pixels are consistently the same across tasks.
Moreover, the average value difference is also minor, 1.3.
Thus, consecutive visual observations are not only similar but also differ only slightly in pixel values.

All the results above show that the effective input difference across consecutive control steps is very small. 
However, existing diffusion-based VLA models still recompute the full DNN for every inference.
Thus, we find that this temporal redundancy is a great opportunity for reducing computation by leveraging prior results and executing only on the changed portions of the input.
\section{Temporal-Aware Bit-Level Sparsity}
\label{sec:sparsity}

\Sect{sec:motiv} shows that, since a robot typically operates at a high frequency, the visual observations of consecutive control steps are highly similar.
However, existing diffusion-based VLAs still perform the full model inference for each control step.
This leads to substantial redundant computation.
In this section, we introduce our \textit{temporal-aware bit-level sparsity} algorithm to leverage this temporal redundancy.
We first introduce the overall idea of our algorithm in \Sect{sec:sparsity:idea}.
We then explain our key design decisions in \Sect{sec:sparsity:decision}.

\subsection{Idea}
\label{sec:sparsity:idea}

Prior bit-sparsity techniques~\cite{chen2024bbs, albericio2017bit, lascorz2019bit, sharma2018bit, lu2021distilling, shi2024bitwave, sharify2019laconic, wei2025phi, judd2016stripes} primarily exploit \emph{bit-level sparsity} in weight parameters. 
In these approaches, the computation cost is reduced by skipping zero-valued bits in model weights during bit-serial computation, as shown in the left side of \Fig{fig:bit_sparsity}. 
However, they do not exploit \textit{temporal redundancy} across consecutive inputs, and therefore fail to eliminate the computation redundancy introduced by consecutive robot control steps. 
\Fig{fig:bit_sparsity} illustrates the key difference between prior bit-sparse execution and our temporal-aware bit-level computation.

\para{Example.}
\Fig{fig:bit_sparsity} uses a toy example to show the computational flow of these two processes. 
We use a linear projection as an example, $\mathbf{y}_t = \mathbf{W}\mathbf{x}_t$,
where $\mathbf{x}_t$ is input activation, $\mathbf{W}$ is model weight, and $\mathbf{y}_t$ is output activation. 
In conventional bit-sparse execution, zero-valued bits in the model weight $\mathbf{W}$ are skipped during computation. 
As illustrated by the sparse-weight example on the left side of \Fig{fig:bit_sparsity}, this can reduce around 50\% of bit-level operations.

However, conventional bit sparsity does not exploit the temporal similarity between adjacent inputs, such as $\mathbf{x}_t$ and $\mathbf{x}_{t-1}$ in the right side of \Fig{fig:bit_sparsity}. 
As a result, it still recomputes the full projection for each control step, even when the input changes only slightly.
In contrast, our method explicitly leverages temporal redundancy. 
The same linear projection can be rewritten as,
\begin{align}
\label{eqn:delta_update}
    \mathbf{y}_t 
    &= \mathbf{W}\mathbf{x}_{t-1} + \mathbf{W}(\mathbf{x}_t - \mathbf{x}_{t-1}) \\
    &= \mathbf{y}_{t-1} + \mathbf{W}\Delta \mathbf{x}_t,
    \quad \text{where} \quad
    \Delta \mathbf{x}_t = \mathbf{x}_t - \mathbf{x}_{t-1}. \nonumber
\end{align}
\Eqn{eqn:delta_update} shows that the full linear projection can be replaced by a differential update. 
Since $\Delta \mathbf{x}_t$ is typically much sparser than $\mathbf{x}_t$ itself, this formulation can significantly reduce the arithmetic workload.

Note that we use \emph{sign-magnitude representation} to represent the differential activation $\Delta \mathbf{x}_t$ rather than two's-complement representation. 
Because we find that sign-magnitude representation further reduces the number of active bits that need to be processed.

\begin{figure}
    \centering
    \includegraphics[width=\columnwidth]{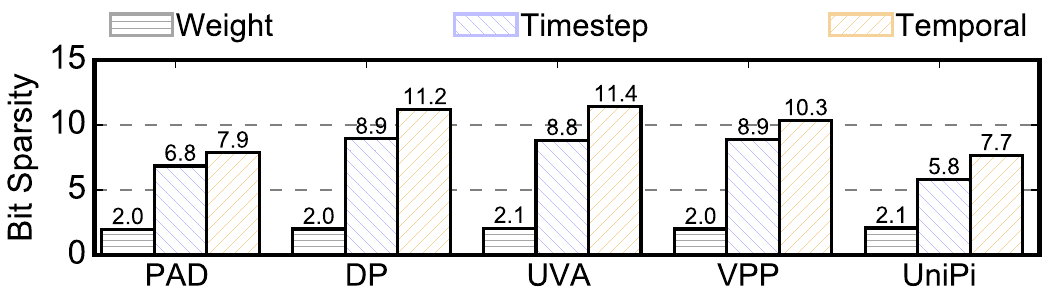}
    \caption{Bit-level sparsity comparison across different diffusion-based VLA models. Bit sparsity is defined as the ratio of the total number of bits to active bits, i.e., ones.}
    \label{fig:sparsity_comparison}
\end{figure}

\Fig{fig:sparsity_comparison} further compares the bit-level sparsity achieved by different methods, where bit sparsity is defined as the ratio of the total number of bits to active bits, i.e., ones.
As shown in \Fig{fig:sparsity_comparison}, the sparsity of model weights alone is much lower than the bit-level sparsity of the activation difference between temporally adjacent inputs. 
For instance, leveraging zero bits in model weights can skip roughly 50\% of the computation.
In contrast, exploiting the temporal redundancy can achieve much higher bit-sparsity and skip over 90\% of the bit-level operations.
This shows that exploiting temporal redundancy can expose more sparse computation opportunities than conventional weight-only bit sparsity.

\para{Method.}
This insight motivates \textit{temporal-aware bit-level sparsity}, which combines temporal differencing with bit-level sparse execution to eliminate redundant computation across consecutive control steps.
Note that, this method is not limited to accelerating the diffusion process; it can also be applied to other modules in a VLA model, such as the visual encoder and multimodal fusion module.
We now describe the general process of our method.

For each DNN module, e.g., linear projection, at the denoising step $t$, we first compute the differential activation, $\Delta \mathbf{x}_t = \mathbf{x}_t - \mathbf{x}_{t-1}$.
Instead of performing full computation, $\mathbf{W}\mathbf{x}_t$, we only compute $\mathbf{W}\Delta \mathbf{x}_t$ and then add it to the cached activation, $\mathbf{y}_{t-1}$, from the previous step, $\mathbf{y}_t = \mathbf{y}_{t-1} + \mathbf{W}\Delta \mathbf{x}_t$.

During this computation, we exploit bit-level sparsity in $\Delta \mathbf{x}_t$.
Each element $\Delta x$ in $\Delta \mathbf{x}_t$ can be represented in binary form, 
$\Delta x = s \sum_{b=0}^{B-1} v_b 2^b$,
where $s$ is the sign bit indicates that whether $\Delta x$ is a negative number.
$B$ is the bit width and $v_b \in \{0,1\}$ indicates whether the $b$-th bit is one.
We then only perform accumulation on active 1 bits.
The differential projection can thus be rewritten as,
\begin{equation}
    \mathbf{W}\Delta \mathbf{x}_t =  (-1)^{\mathbf{s}_{t}} \times \sum_{b=0}^{B-1} \mathbf{W}\times \mathbf{v}_{t,b} \times 2^b,
\end{equation}
where $\mathbf{v}_{t,b}$ is the zero indicator for the $b$-th bit of $\Delta \mathbf{x}_t$. Only the bits with nonzero $\mathbf{v}_{t,b}$ need to be processed; those with zero $\mathbf{v}_{t,b}$ are skipped entirely.
This differential execution can be applied to the dominant linear operators in diffusion-based VLA models, including input projections, MLP projections, Q/K/V projections, cross-attention projections, and output projections.
We discuss how to handle self-attention in \Sect{sec:sparsity:decision}.

\subsection{Design Decisions}
\label{sec:sparsity:decision}

Next, we explain some of the key design decisions in our algorithm.

\para{Why Frame-wise Similarity?}
We have witnessed a broad set of studies that exploit similarity across denoising steps~\cite{zhao2024real, zou2024accelerating, chen2024delta, ma2024deepcache, wimbauer2024cache, liu2024smoothcache}. 
Thus, a natural question is why we focus on \textit{frame-wise similarity across control steps} rather than \textit{similarity across denoising steps}. 
Here, we provide two key reasons.

First, our experiments show that frame-wise similarity introduces higher bit-level sparsity than step-wise similarity. 
In \Fig{fig:sparsity_comparison}, we compare the bit sparsity achieved by these two techniques: differential computing across consecutive robot observations and differential computing across adjacent denoising steps. 
The results show that exploiting frame-wise similarity consistently achieves higher bit sparsity, i.e., more bit-level computation to be skipped, than exploiting step similarity. 
Moreover, as robotic systems move toward higher control frequency, we expect frame-wise similarity between adjacent control steps to become even stronger.

Second, leveraging frame-wise similarity is more robust to future model trends. 
While many prior works~\cite{heo2025exion, zhao2024real, kim2025ditto, liu2025astraea} exploit similarity across denoising steps, this source of similarity may diminish as diffusion models increasingly apply distillation techniques to reduce the number of denoising steps~\cite{hacohen2024LTXVideo, zhang2025turbodiffusion, li2025unleashing, salimans2022progressive}. 
As the number of steps becomes smaller, the difference between adjacent steps generally becomes larger.
Thus, no step-level similarity can be exploited.
On the other hand, frame-wise similarity across action steps will persist, since future robotic systems will continue to demand high control frequency. 
Therefore, frame-wise similarity is naturally a more long-term optimization target.

Nevertheless, these two forms of similarity are \textit{orthogonal}. 
Step-level bit sparsity reduces redundancy within a single diffusion process, whereas temporal-aware bit-level sparsity eliminates redundancy across consecutive robot actions. 
In principle, the two techniques can be applied together. 
% In \Sect{??}, we will also report results that leverage both forms of redundancy.

\begin{figure}[t]
\centering
\begin{minipage}[t]{0.48\columnwidth}
  \centering
  \includegraphics[width=\columnwidth]{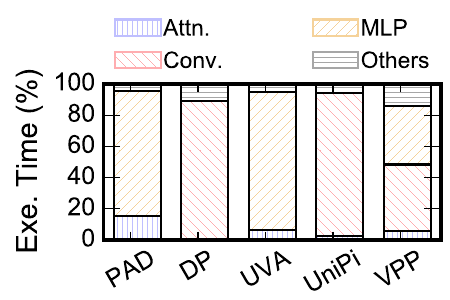}
  \caption{The execution breakdown of VLA models. Self-attention does not dominate the execution.}
  \label{fig:exec_breakdown_attn}
\end{minipage}
\hspace{2pt}
\begin{minipage}[t]{0.48\columnwidth}
  \centering
  \includegraphics[width=\columnwidth]{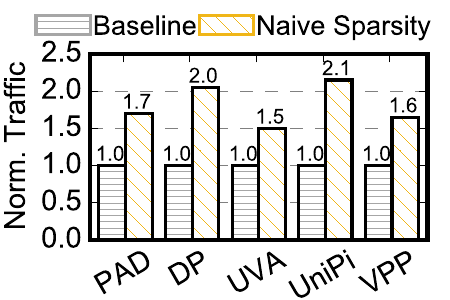}
  \caption{The off-chip data traffic comparison between normal execution and naive bit-serial computation.}
  \label{fig:dram_traffic}
\end{minipage}
\end{figure}

\para{Handling Self-Attention.}
Our temporal-aware bit-level sparsity described in \Sect{sec:sparsity:idea} naturally applies to linear projections, where only one operand varies across action controls while the other operand (the model weight) remains constant.
However, self-attention introduces an additional challenge: both the query and key values vary across inputs, making differential computation ineffective.
The attention score $\mathbf{Q}_t \mathbf{K}_t^{T}$ have to be expanded as,
\begin{equation}
    \mathbf{Q}_t \mathbf{K}_t^{T}
    =
    \mathbf{Q}_{t-1}\mathbf{K}_{t-1}^{T}
    + \Delta \mathbf{Q}_t \mathbf{K}_{t-1}^{T}
    + \mathbf{Q}_{t}\Delta \mathbf{K}_t^{T},
\end{equation}
where $\mathbf{Q}_{t-1}$ and $\mathbf{K}_{t-1}$ denote the query and key tensors from the previous control step, respectively, and
\begin{equation}
    \Delta \mathbf{Q}_t = \mathbf{Q}_t - \mathbf{Q}_{t-1}, \quad \text{and} \quad
    \Delta \mathbf{K}_t = \mathbf{K}_t - \mathbf{K}_{t-1}
\end{equation}
represent the corresponding differential updates.

Compared to linear projections, this expansion introduces multiple additions, making it less efficient.
Therefore, for simplicity, we apply only conventional bit-level sparsity to self-attention, without leveraging temporal similarity.
In other words, during bit-sparsity execution, we only skip zero bits from one operand (e.g., $\mathbf{K}$), while computing the attention score directly on the current activations.
This design simplifies the attention computation while still reducing the number of required bit-level additions.

Meanwhile, our execution breakdown across different diffusion-based VLA models shows that self-attention accounts for only a small fraction, 5.9\%, of the total execution, as shown in \Fig{fig:exec_breakdown_attn}.
As a result, this limitation has a minimal performance impact, while accelerating these linear projections with temporal-aware bit-level sparsity captures most of the performance benefit.
\section{Speculative Inference}
\label{sec:spec}

\begin{figure}
    \centering
    \includegraphics[width=\columnwidth]{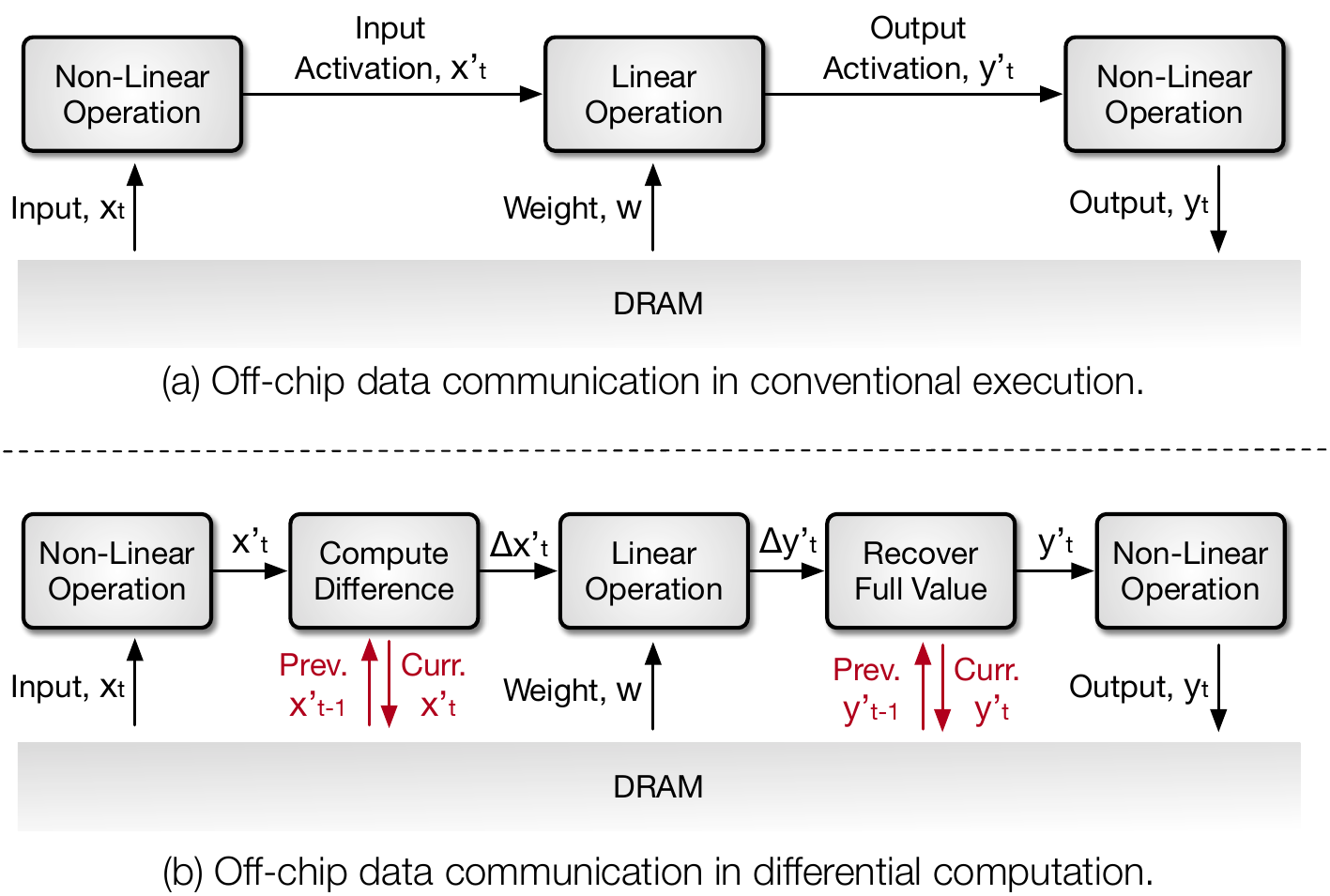}
    \caption{Comparison of off-chip data traffic in (a) conventional execution and (b) naively applying our temporal-aware bit sparsity. In this naive sparse execution, extra data traffic is required to accommodate non-linear operations.}
    \label{fig:data_traffic_issues}
\end{figure}

\para{Issues.}
While temporal-aware bit-level sparsity can effectively reduce the total number of bit-level operations, it also introduces a new challenge: \textit{extra off-chip data traffic}.
\Fig{fig:data_traffic_issues} gives a simple example in which a linear operation is sandwiched between two nonlinear operations.
For simplicity, we assume that the on-chip buffer is large enough to hold all required intermediate values.

In conventional inference, each layer only needs to fetch the current input activation, $\mathbf{x}'_t$, and the model weights, $\mathbf{W}$, into the on-chip buffer, as shown in \Fig{fig:data_traffic_issues}a.
After computation, the output activation $\mathbf{y}'_t$ is then written back to off-chip memory.

If no nonlinear operation is involved, our temporal-aware bit-sparse execution can follow a similar communication pattern, except that it reads and writes the differential activations, $\Delta \mathbf{x}'_t$ and $\Delta \mathbf{y}'_t$, instead of the full activations, $\mathbf{x}'_t$ and $\mathbf{y}'_t$.

However, once nonlinear operators are introduced, to preserve correctness, the differential results must be reconstructed into full-value activations before applying the nonlinear operation.
This requires loading the previous intermediate results, e.g., $\mathbf{x}'_{t-1}$ and $\mathbf{y}'_{t-1}$, to recover the current full-value activations, e.g., $\mathbf{x}'_t$ and $\mathbf{y}'_t$.
Moreover, these reconstructed activations must also be preserved so that subsequent denoising steps can be computed correctly.
As a result, our temporal-aware bit-sparsity can increase the amount of intermediate data that must be read from and written to memory during inference.
\Fig{fig:dram_traffic} shows the comparison of the off-chip data communication between the conventional execution and naively applying our idea.
Overall, naively applying our temporal-aware bit sparsity would increase 80\% of the total data traffic.

\begin{figure}
    \centering
    \includegraphics[width=\columnwidth]{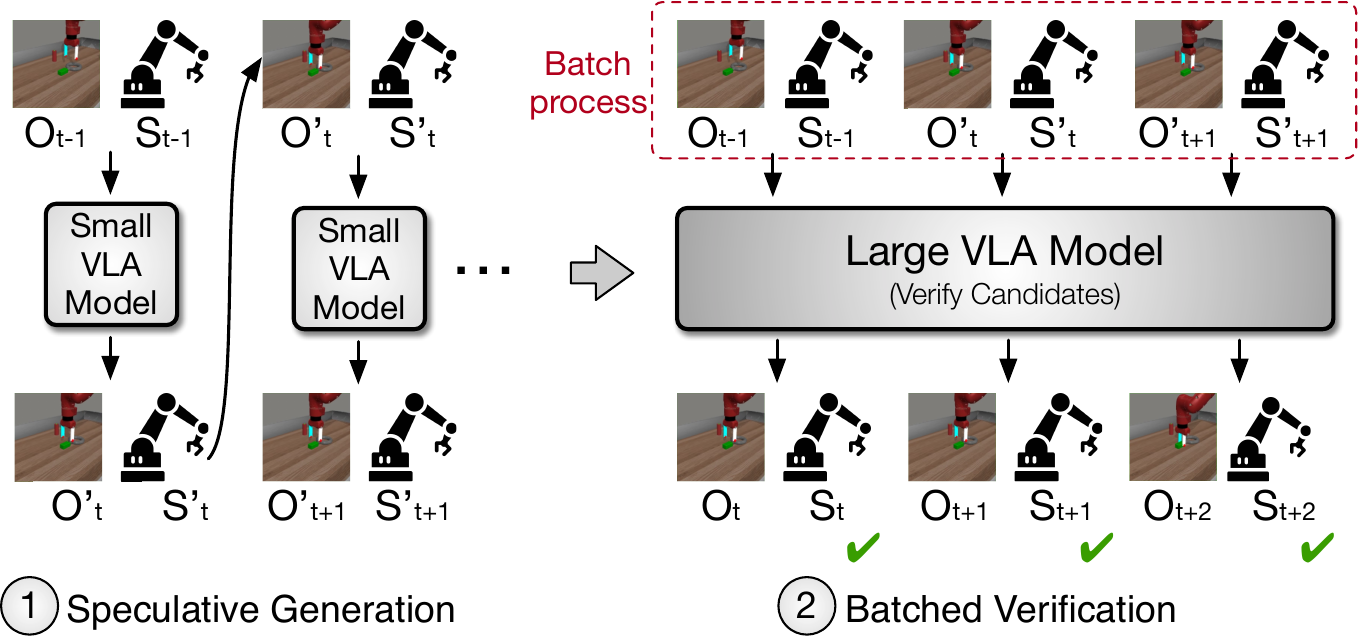}
    \caption{Overview of our technique, \textit{speculative inference}. It first allows a lightweight VLA model to generate a sequence of speculative candidates for future steps. These candidates are then verified together by a more accurate VLA model in a batched manner. Thus, we can amortize the data traffic of weight and activation accesses across verified candidates.}
    \label{fig:speculative_inference}
\end{figure}

\para{Idea.}
To reduce the off-chip data traffic, we propose a technique called \textit{speculative inference}, as shown in \Fig{fig:speculative_inference}. 
Instead of generating robot actions one step at a time using only the original diffusion-based VLA model, we first use a \textit{small VLA model} to generate a sequence of speculative candidates for multiple future control steps. 
Starting from the current observation and robot state, the small model recursively predicts the next observation and state.
Note that, these speculative candidates cannot be directly used to control the robot. 
Thus, during speculative rollout, the inputs to the small model are the previously predicted observations and states, e.g., $\mathbf{O}'_{t}$ and $\mathbf{S}'_{t}$.
In this way, we produce a speculative trajectory,
$$
\{(\mathbf{O}'_{t}, \mathbf{S}'_{t}), (\mathbf{O}'_{t+1}, \mathbf{S}'_{t+1}), \dots, (\mathbf{O}'_{t+k-1}, \mathbf{S}'_{t+k-1})\},
$$
where $k$ is the speculative window size.

Because the small model is lightweight, the cost of candidate generation is modest. 
Meanwhile, our temporal-aware bit-level sparsity described in \Sect{sec:sparsity:idea} can also be applied to the small model, further reducing its computational overhead. 

However, because the small model is less accurate, its predictions cannot be executed directly. 
Therefore, in the second stage, a more accurate \emph{large VLA model} verifies the speculative candidates in a batched manner, as shown in \Fig{fig:speculative_inference}. 
This batched verification is the key to reducing data communication. 
Because multiple speculative candidates are verified together, the weights of the large model need to be loaded only once for several control steps, rather than being repeatedly fetched in a frame-by-frame fashion. 
Similarly, the activations of multiple speculative candidates can be loaded and processed together.
Therefore, we can amortize both weight traffic and activation traffic across the verification window.
As a result, we reduce the overall data movement of VLA inference.

\begin{figure}
    \centering
    \includegraphics[width=\columnwidth]{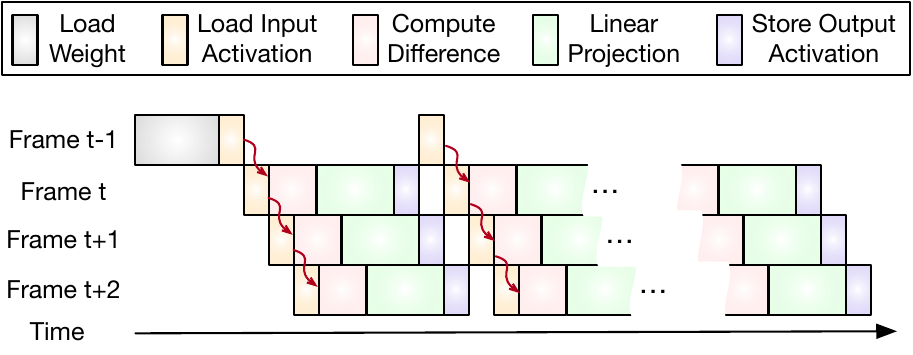}
    \caption{Illustration of batched verification in speculative inference. Multiple speculative frames are verified together by the large VLA model, allowing the same model weights and activations to be loaded once and reused multiple times.}
    \label{fig:batch_verification}
\end{figure}

\para{Batched Verification.}
\Fig{fig:batch_verification} further gives an example of executing a linear projection module using \textit{batched verification}. 
Instead of verifying each speculative candidate independently, the large VLA model verifies multiple future frames together in a batched manner. 
Here, to accommodate the limited on-chip buffer, the input activations are loaded piece by piece.
The input activations of several speculative frames are processed in an interleaved fashion by the same operator pipeline, i.e., a linear projection module.

Overall, batched verification enables reusing two major sources of data. 
First, the model weights only need to be loaded once and can then be reused across all frames within the speculative window.
Here, the window size is 3. 
% In conventional frame-by-frame execution, each frame must repeatedly fetch the same model weights from off-chip memory before performing computation. 
% In contrast, 
This way, batched verification amortizes the weight-loading cost.
Second, batched verification also enables input activation reuse. 
Since the speculative candidates of multiple future frames are available in advance, their input activations can be loaded and processed together. 
For example, the activations of frame $t$ and frame $t+1$ can be jointly reused when computing the corresponding differential activations, $\Delta \mathbf{x}_t$ and $\Delta \mathbf{x}_{t+1}$. 
The red lines in \Fig{fig:batch_verification} show the data dependencies of \textit{compute difference} stage.
% As a result, batched verification amortizes both weight and activation loading, largely reducing off-chip data communication.

\hl{
After the large VLA model produces the verified outputs, e.g., $\mathbf{S}_{t+1}$ and $\mathbf{S}_{t+2}$, we compare them against the corresponding predictions from the small VLA model, i.e., $\mathbf{S}'_{t+1}$ and $\mathbf{S}'_{t+2}$, sequentially. 
If the prediction error,
$\Delta \mathbf{S}_{t+1} = |\mathbf{S}_{t+1} - \mathbf{S}'_{t+1}|$,
is smaller than a predefined threshold, $\theta_{th}$, then the verified output $\mathbf{S}_{t+1}$ is accepted and can be used to control the robot. 
Here, we set the predefined threshold to be 0.01. 
\mbox{\Sect{sec:eval:sen}} further shows the performance and accuracy sensitivity to the threshold, $\theta_{th}$.
If the error $\Delta \mathbf{S}_{t+1}$ exceeds the threshold $\theta_{th}$, we stop at the step $t+1$.
The action $\mathbf{S}_{t+1}$ and all subsequent speculative actions in the window are discarded.
The next round of speculation restarts from the step $t+1$.
}

In summary, speculative inference mitigates the additional off-chip data traffic introduced by differential computing. 
By using a lightweight small VLA model to generate speculative candidates and a large VLA model to verify them in batches, our design amortizes both weight and activation loading across multiple future control steps, largely reducing overall off-chip data traffic.

\section{\proj Architecture}
\label{sec:arch}

Although a wide range of bit-sparsity accelerators have been proposed~\cite{chen2024bbs, albericio2017bit, lascorz2019bit, sharma2018bit, lu2021distilling, shi2024bitwave, sharify2019laconic, wei2025phi, judd2016stripes}, existing designs are not tailored to support our temporal-aware bit-level sparsity. 
To fully exploit this new form of sparsity, we propose a novel hardware architecture, \proj, that is co-designed with our algorithms. 
We first provide an overview of the architecture in \Sect{sec:arch:overview}, and then present the key architectural components in \Sect{sec:arch:pe} and \Sect{sec:arch:encoder}.

\subsection{Overview}
\label{sec:arch:overview}

\begin{figure}
    \centering
    \includegraphics[width=\columnwidth]{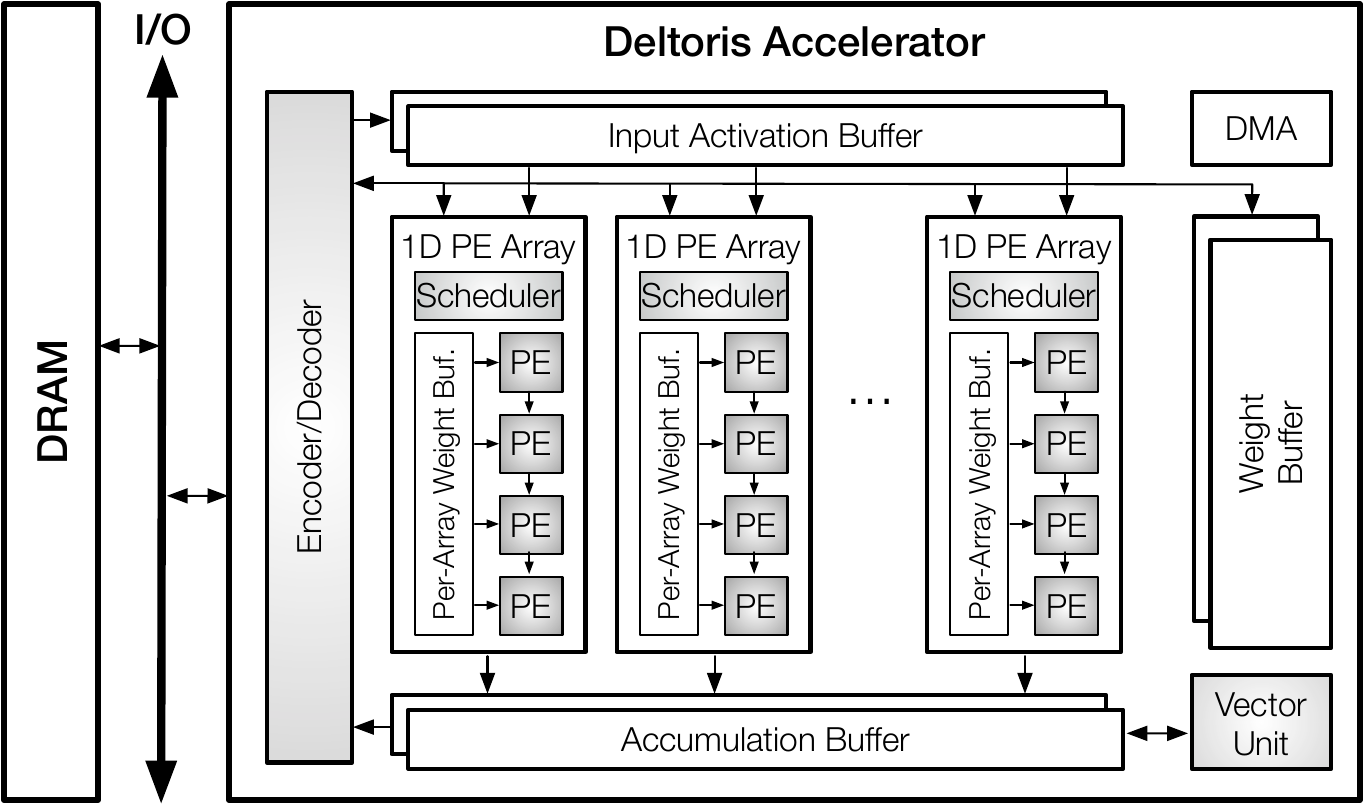}
    \caption{Overall architecture of the \proj. The design consists of multiple 1D PE arrays, a vector unit, an input activation buffer, a weight buffer, and an accumulation buffer. All three buffers are double-buffered to overlap compute with data communication. The 1D PE arrays are designed for bit-serial matrix multiplications, while the vector unit handles nonlinear operations. The encoder/decoder compresses data before it is read from or written back to DRAM.}
    \label{fig:arch}
\end{figure}

\Fig{fig:arch} shows the overall architecture of the \proj accelerator.
At a high level, the design integrates an on-chip buffer, an encoder/decoder, a vector unit, and a set of independent 1D PE arrays to support our temporal-aware bit-sparse algorithm.

The on-chip buffer is partitioned into an input activation buffer, a global weight buffer, and an accumulation buffer. 
First, input activations are fetched from off-chip DRAM into the input activation buffer via the DMA engine, while model weights are first buffered in the global weight buffer and then broadcast to per-array local weight buffers.
After computation, the results are written to the accumulation buffer.
Note that, all on-chip buffers are double-buffered to overlap computation with data communication.
Furthermore, an encoder/decoder module compresses all data before it is transferred to or from global memory to reduce off-chip traffic.
The design of the encoder/decoder is explained in \Sect{sec:arch:encoder}.

The core of \proj is a set of 1D PE arrays designed to support our temporal-aware bit-sparse algorithm.
Each PE array consists of a local weight buffer with a set of PEs targeted for bit-serial matrix multiplications.
We further explain the design of PE array in \Sect{sec:arch:pe}.
Note that, our PE array design eliminates the workload imbalance that commonly exists in conventional bit-serial accelerators.
Lastly, the output results are written to the accumulation buffer and then either written back to the off-chip memory or forwarded to the vector unit for element-wise or nonlinear operations.

To maximize throughput, \proj also adopts a pipelined execution that overlaps data loading, decoding, computation, encoding, and write-back by leveraging the double-buffered buffers. 
Thus, we can hide the execution of different stages.

\hl{
To support speculative inference, the PE array executes both the small draft model and the large verification model using the same bit-serial PE array. 
Both models are executed via our temporal-aware bit-sparsity algorithm described in \mbox{\Sect{sec:sparsity}}.
The only difference is that the small draft model does not leverage the batch verification to optimize the off-chip data traffic.
The generated candidates and verification outputs are stored in on-chip accumulation buffers, and the error comparison is performed on the vector unit.. 
The control logic is managed by the microcontroller on the accelerator. 
Here, we do not explicitly show it in \mbox{\Fig{fig:arch}}.
}

\subsection{PE Array}
\label{sec:arch:pe}

\begin{figure*}
    \centering
    \includegraphics[width=\textwidth]{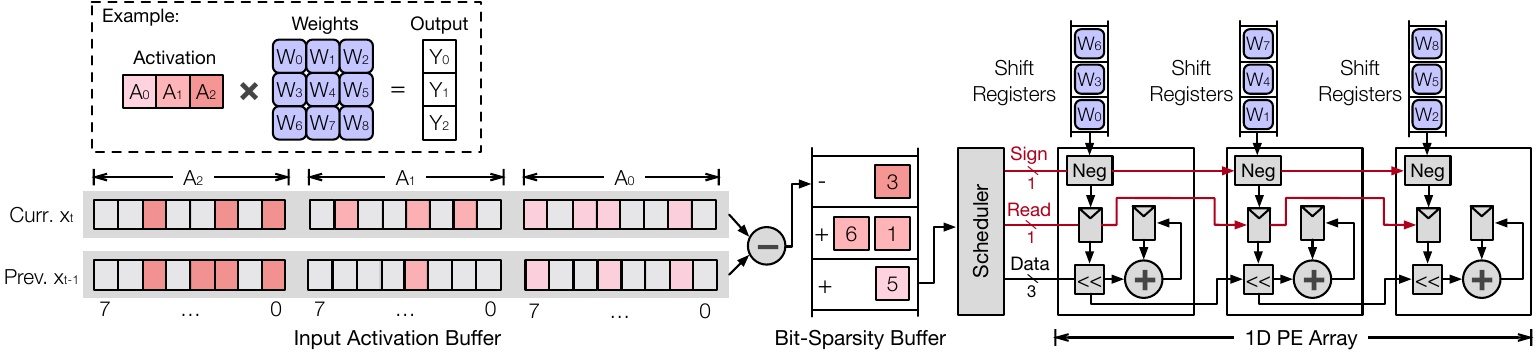}
    \caption{Design of 1D PE array. Here, we show how a $1 \times 3$ activation and a $3\times3$ weight are mapped to the PE array. We adopt an output-stationary dataflow. We first compute the difference between the current and previous activations to obtain a sparse differential activation, and store only the active bits in the bit-sparsity buffer. Then, the scheduler issues these sparse bit events to the 1D PE array, determines when to read the next weight, and specifies the required bit shifts for accumulation. Weights are delivered through local shift registers. Note that, all PEs in the same row share common control signals, such as read and sign (highlighted by red lines). Thus, work imbalance has been eliminated.}
    \label{fig:pe_array}
\end{figure*}

\Fig{fig:pe_array} shows the design of our 1D PE array. 
Note that, our PE array can accelerate all linear modules in VLA models, e.g., the visual encoder and the action prediction in \Fig{fig:vla_workflow}.
Here, we just use a toy example to show how our PE array computes a linear projection.
For simplicity, we consider a small matrix multiplication between a $1\times 3$ activation and a $3\times 3$ weight.

Our computation starts with reading the current activation and the previous activation from the input activation buffer.
We first compute the activation difference between the two consecutive inputs and identify the active bits, i.e., ``1''s, that must be processed.
Only the bit indices of these active bits are stored in the bit-sparsity buffer.
The bit-sparsity buffer records the information required for execution, including the sign of activation differences and the bit indices of active bits.
For instance, in activation $A_0$, there is only one active bit, i.e., the 5th bit, that needs to be computed. We thus store this bit index and the sign bit to the bit sparsity buffer.

\para{Scheduler.}
These sparse bits are then dispatched by the scheduler to the 1D PE array.
The scheduler orchestrates the computation between sparse activation bits and model weights.
At each cycle, the scheduler determines three pieces of information:
1) the shift amount required for the current weight value,
2) whether the PE array should read the next weight value, and
3) whether the current weight should be added to or subtracted from the accumulated sum.

The shift amount is determined by the bit index, as shown by the data line in \Fig{fig:pe_array}.
To support weight update and sign control, we introduce two additional signals: \textit{read} and \textit{sign}.
These signals are propagated through the PE array, similar to the data line.
Because all PEs in the same row operate under a shared control flow, we can avoid the need for dedicated per-PE control signals.

\para{PE Design.}
The core computation units of \proj are a set of 1D PE arrays.
The design is conceptually similar to a systolic array.
Each 1D PE array consists of a sequence of PEs connected along one dimension.
E.g., from left to right in \Fig{fig:pe_array}.
We adopt an \textit{output-stationary} dataflow, in which the accumulated output activations remain in the PE registers while the sparse activation bits stream through the array.
Meanwhile, each PE has its own local shift registers to read its model weights.

Within the array, each PE is responsible for computing one element of the output activation. 
The activation bit indices are forwarded along the row, while the corresponding weights are fed to the PEs according to the read signal. 
In the example shown in \Fig{fig:pe_array}, during the first cycle, the left-most PE receives bit index $5$, reads weight $W_0$, and performs the corresponding bit-level accumulation. 
In the next cycle, this PE forwards the bit index $5$ to the middle PE, together with the read and sign signals. 
The middle PE then reads weight $W_1$ and performs its own computation, $2^5 \times W_1$. 
Meanwhile, the left-most PE advances to the next active bit, reads weight $W_3$, and starts computing $2^1 \times W_3$.
This process continues until all output activations assigned to the PE array have been fully computed.
Once the array is fully pipelined, all PEs in the same row remain fully utilized.
Thus, we eliminate the workload imbalance commonly observed in conventional bit-sparsity accelerators.

\para{Weight Sharing.}
To minimize on-chip SRAM read traffic, \proj adopts a weight sharing technique across multiple 1D PE arrays.
Rather than allowing each PE array to fetch weights individually, which leads to redundant SRAM accesses, we group $M$ PE-array rows and share weight reads across them.
$M$ is set to 16 in our case.
Each time, a batch of weights is fetched once from SRAM and broadcast to all $M$ rows.
Meanwhile, these $M$ rows of PE arrays continue processing the previously fetched weight batch.
After $M$ rows complete execution, the next batch of weights is fetched and broadcast.
This way, we amortize SRAM accesses across multiple PE arrays and reduce on-chip data movement.

% Overall, our PE array provides an efficient execution substrate for temporal-aware bit-sparsity. 
% With its weight-stationary dataflow and shared row-wise control, the design reduces compute and control overhead while maintaining high PE utilization.

\subsection{Encoder and Decoder}
\label{sec:arch:encoder}

To further reduce off-chip traffic, \proj designs a lightweight encoder/decoder module that compresses data before it is written back to DRAM. 
Our encoder design leverages the temporal similarity of intermediate data across consecutive frames. 
Specifically, we compress the first frame and model weights using vector quantization~\cite{linde1980algorithm} to achieve high compression efficiency, while subsequent frames are encoded via a lightweight compressed sparse row (CSR) format to capture the sparse differences between frames. 
This hybrid compression scheme can further reduce the data communication overhead of intermediate data and model weights. 
Both compression and decompression are pipelined with computation and data fetching.
Thus, it does not increase the overall latency.

\section{Experimental Setup}
\label{sec:exp}

\para{Hardware Setup.} 
We develop validated RTL implementations for the \proj hardware. 
The hardware design is based on a 1D systolic array architecture in \Sect{sec:arch:pe}.
\proj consists of 128 1D PE arrays.
Each array consists of 64 PEs clocked at 1~GHz.
Each PE has one shift-based accumulator and 64~B shift registers to store local weights.
Meanwhile, the vector unit in \proj is clocked at 250~MHz, which consists of 16 parallel lanes, each capable of performing element-wise operations and nonlinear activation functions.
The encoder/decoder has 8 units to perform compression/decompression in parallel.
The design has 2~MB of on-chip SRAM in total, partitioned into the input activation buffer, weight buffer, and accumulation buffer, with a bank granularity of 64~KB.

\para{Experimental Methodology.} 
We develop the RTL implementation of \proj using Synopsys synthesis tools and Cadence layout design tools in TSMC 16~nm FinFET technology, with SRAM generated by an Arm memory compiler.
Power is simulated using Synopsys PrimeTimePX with fully annotated switching activity.
To enable a fair comparison against the mobile GPU in the Nvidia Orin SoC, we scale the results to the 8~nm process node using DeepScaleTool~\cite{stillmaker2017scaling, sarangi2021deepscaletool}. 
The off-chip memory is modeled using four Micron 32~Gb LPDDR4-3200 channels~\cite{lpddr4}, and DRAM energy is estimated using Micron’s System Power Calculator~\cite{microdrampower}. 
To evaluate system-level performance, we build a cycle-accurate simulator that captures component-level latency and power consumption.
The reported energy and performance of \proj are derived from post-layout RTL measurements and normalized to the 8~nm node.

\para{Area Overhead.}
\proj has comparable area compared to the prior bit-sparsity accelerators~\cite{albericio2017bit, chen2024bbs} and diffusion accelerators~\cite{heo2025exion, kong2024cambricon, kim2025ditto}.
The total area of our design is 2.45 mm$^2$, which is comparable to other accelerators in \Tbl{tab:hw}.
All PE arrays take 22\% of the total area; the SRAMs comprise the majority of the remaining area.
The detailed component-level configuration is shown in \Tbl{tab:hw}.

\begin{table} 
\caption{\hl{Evaluation benchmarks.}}
\resizebox{0.9\columnwidth}{!}{
\renewcommand*{\arraystretch}{1}
\renewcommand*{\tabcolsep}{10pt}
\begin{tabular}{ cccc } 
\toprule[0.15em]
\textbf{\specialcell{Algorithm}} & \textbf{Dataset} & \textbf{Metric} & \textbf{Venue} \\ 
\midrule[0.05em]
PAD~\cite{guo2024prediction} & Meta-World~\cite{mclean2025metaworld} & Success Rate & NeurIPS 2024 \\
\midrule[0.05em]
DP~\cite{chi2023diffusion} & PushT~\cite{gym-pusht} & Success Rate & RSS 2023 \\
\midrule[0.05em]
UVA~\cite{li2025unified} & \specialcell{PushT~\cite{gym-pusht}\\LIBERO~\cite{liu2023libero}} & Success Rate & RSS 2025 \\
% DP~\cite{chi2023diffusion} & ?? & Success Rate & RSS'23 \\
% UVA~\cite{li2025unified} & ?? & Success Rate & RSS'25 \\

\bottomrule[0.15em]
\end{tabular}
}
\label{tab:sw}
\end{table}

\para{Software Baselines.}
\hl{In \mbox{\Tbl{tab:sw}}, we show our software setup.
We evaluate \mbox{\proj} using three representative diffusion-based VLA algorithms: PAD~\mbox{\cite{guo2024prediction}}, DP~\mbox{\cite{chi2023diffusion}}, and UVA~\mbox{\cite{li2025unified}}.
To evaluate the robustness of our techniques, we adopt three widely used VLA datasets: Meta-World~\mbox{\cite{mclean2025metaworld}}, PushT~\mbox{\cite{gym-pusht}}, and LIBERO~\mbox{\cite{liu2023libero}}.
We use the success rate of task completion as the accuracy evaluation metric.}

In speculative inference (\Sect{sec:spec}), it requires a small model and a large model.
However, the inputs of existing diffusion-based VLA models are not directly compatible.
To address this issue, we adopt a distillation approach similar to prior work~\cite{liu2025astraea, li2023autodiffusion} to derive a lightweight model from the original diffusion model.
We then use the distilled model as the small model and the original model as the large model.
\hl{
Specifically, we do not train a separate small VLA model from scratch. 
Instead, the draft model uses the same network architecture and model dimensions as the original diffusion-based VLA model, but runs with a reduced number of denoising timesteps. 
In our evaluation, the draft model uses roughly 1/5 of the denoising computation of the original model.

We adopt this training-free distillation method for two reasons. First, our goal is not to propose a new VLA model architecture or a new distillation/training method, but to study how speculative inference can reduce the cost of diffusion-based VLA execution. Second, training a separate small model would introduce additional training overhead and may require extra alignment between the draft and large models, if their architectures or embeddings differ.

}

\begin{table} 
\caption{Hardware baselines. The area is reported in 16 nm.}
\resizebox{\columnwidth}{!}{
\renewcommand*{\arraystretch}{1}
\renewcommand*{\tabcolsep}{3pt}
\begin{tabular}{ cccc } 
\toprule[0.15em]
\textbf{\specialcell{Baselines}} & \textbf{Configuration} &  \textbf{Area (mm$^2$)} \\ 
\midrule[0.05em]
\mode{Pragmatic}~\cite{albericio2017bit} & 512 inner product units &  2.26 \\
\mode{BBS}~\cite{chen2024bbs} & 32$\times$32 BitVert Units &  2.50 \\
\mode{Cambricon-D}~\cite{kong2024cambricon} & 32 PEs, each with 4 A8W8 MACs and 60 A4W8 MACs  & 2.41 \\
\mode{Exion}~\cite{heo2025exion} & 64 DPU units  & 2.39 \\
\mode{Ditto}~\cite{kim2025ditto} & 32$\times$16 compute units, each with 4 A4W8 MACs & 2.32 \\
\mode{\proj} & 128 rows of PE array with 64 PEs & 2.45 \\
\bottomrule[0.15em]
\end{tabular}
}
\label{tab:hw}
\end{table}

\para{Hardware Baselines.}
We compare \proj against six hardware baselines.
Specifically, we compare two bit-sparsity accelerators and three diffusion-based accelerators:
\begin{itemize}
    \item \mode{GPU}: a mobile Ampere GPU in Nvidia Jetson Orin SoC~\cite{orinsoc}.
    \item \mode{Pragmatic}~\cite{albericio2017bit}: a bit-serial accelerator that supports irregular bit sparsity computation.
    \item \mode{BBS}~\cite{chen2024bbs}: an algorithm-hardware co-design that supports bit-level sparsity with some algorithmic approximations.
    \item \mode{Cambricon-D}~\cite{kong2024cambricon}: a mix-precision diffusion-specific accelerator that leverages the timestep-level similarity to perform differential computing.
    \item \mode{Exion}~\cite{heo2025exion}: a diffusion accelerator that leverages the attention sparsity and inter-timestep similarity.
    \item \mode{Ditto}~\cite{kim2025ditto}: similar to \mode{Cambricon-D}, a mix-precision sparse accelerator that leverages the timestep-level similarity.
\end{itemize}
To ensure a fair comparison, we configure the compute throughput of the PE arrays across all accelerators to be equivalent to $ 32 \times 32$ 8-bit MAC units. 
We also configure all designs with the same on-chip buffer capacity of 2~MB. 
Meanwhile, the overall energy consumption of each accelerator design is also scaled to the 8~nm technology node, enabling a fair comparison against \mode{GPU}.
\Tbl{tab:hw} shows the detailed configurations of hardware baselines.

% \para{Variants.}
% To dissect our contributions in algorithm and architecture, we evaluate five variants of \proj to separate the contributions proposed in our paper:
% \begin{itemize}
% \item \mode{\proj-sw}: this variant is a GPU implementation of our technique, Gaussian freezing.
% \item \mode{\proj-hw}: this variant is a full-fledged implementation with our hardware support.
% \end{itemize}

\section{Evaluation}
\label{sec:eval}

% We first demonstrate that \proj achieves comparable accuracy to the baseline (\Sect{sec:eval:acc}). 
% We then show the speedup and energy savings of \proj compared to prior hardware baselines (\Sect{sec:eval:perf}).
% We next dissect the performance contribution of \proj in \Sect{sec:eval:abl}.
% Lastly, we perform a sensitivity study to understand \proj's performance gains under different settings (\Sect{sec:eval:sen}). 

\subsection{Accuracy}
\label{sec:eval:acc}

% \begin{figure}
%     \centering
%     \includegraphics[width=\columnwidth]{accruacy}
%     \caption{Accuracy comparison between the baseline algorithm and our proposed algorithm. Our algorithmic optimizations introduce negligible accuracy loss.}
%     \label{fig:acc}
% \end{figure}

\begin{figure}[t]
\centering
\subfloat[Success rate. Higher is better.]{
    \label{fig:success_rate}
    \includegraphics[width=0.49\columnwidth]{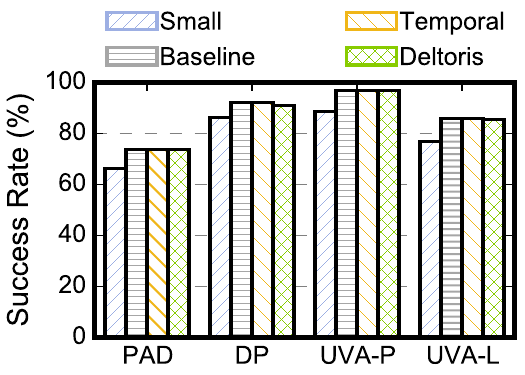}}
\subfloat[Qualitative result.]{
    \label{fig:result_example}
    \includegraphics[width=0.4\columnwidth]{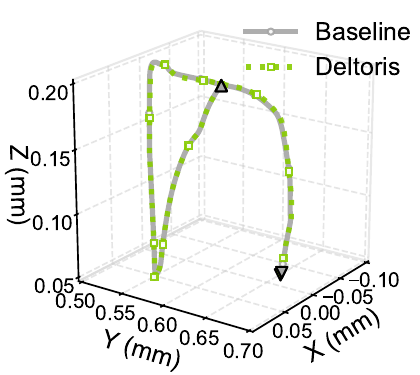}}
\caption{\hl{Accuracy comparison between the baseline algorithm and our proposed algorithm. Our algorithmic optimizations introduce negligible accuracy loss. UVA-P: UVA result on PushT dataset; UVA-L: UVA result on LIBERO dataset.}}
\label{fig:acc}
\end{figure}

\begin{figure}[t]
\centering
\subfloat[\hl{Speedup evaluation. Higher is better.}]{
    \label{fig:speedup}
    \includegraphics[width=\columnwidth]{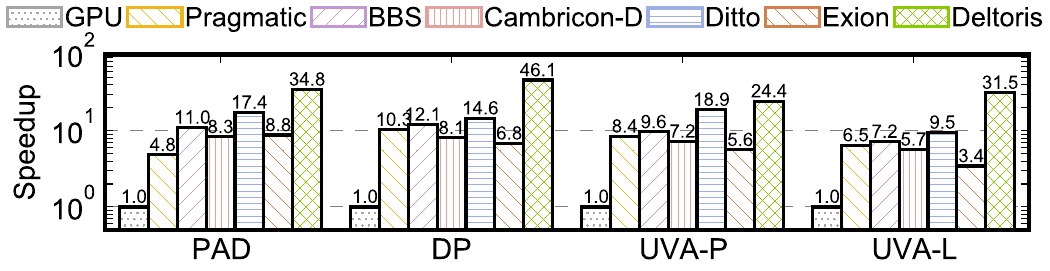}}
\\
\subfloat[\hl{Energy savings. Higher is better.}]{
    \label{fig:energy}
    \includegraphics[width=\columnwidth]{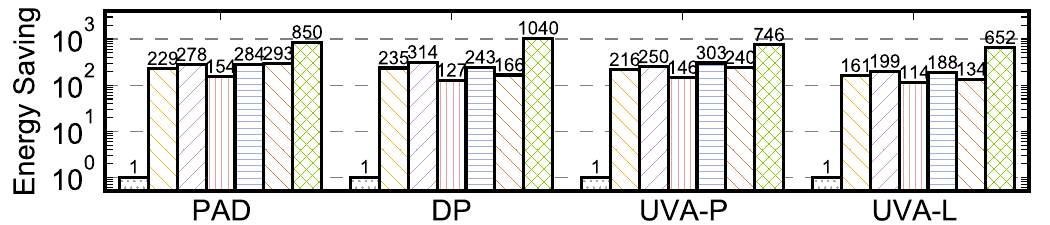}}
\\
\subfloat[\hl{Overall normalized operation counts. Lower is better.}]{
    \label{fig:ops}
    \includegraphics[width=\columnwidth]{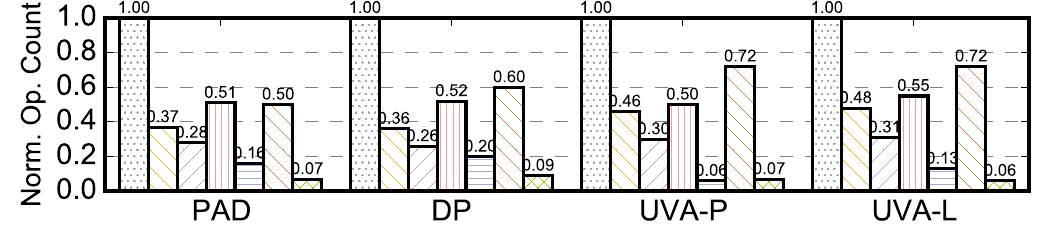}}
\caption{\hl{Overall performance evaluation against prior work. Numbers are normalized to \mbox{\mode{GPU}}. UVA-P: UVA result on PushT dataset; UVA-L: UVA result on LIBERO dataset.}}
\label{fig:perf}
\end{figure}

We first evaluate the accuracy impact of our proposed techniques: temporal-aware bit sparsity and speculative inference.
As shown in \Fig{fig:acc}, temporal-aware bit sparsity, \mode{Temporal}, is \textit{bit-accurate}.
It matches the baseline exactly.
When combined with speculative inference, \mode{\proj} incurs only negligible accuracy loss across all evaluated algorithms.
\hl{On average, \mbox{\mode{\proj}} introduces 0.2\% accuracy loss, which is negligible.
In comparison, only applying small models, \mbox{\mode{Small}}, shows noticeable accuracy loss ($>$ 7.7\%).}
These results indicate that speculative inference is sufficiently robust for diverse robotic tasks and cannot be simply replaced by a small VLA model.
\Fig{fig:result_example} further shows a qualitative comparison.
% between \proj and baselines.

\subsection{Performance}
\label{sec:eval:perf}

\para{Speedup.}
\hl{\mbox{\Fig{fig:speedup}} shows the performance comparison across different accelerators.
Overall, \mbox{\mode{\proj}} achieves the highest speedup across all evaluated algorithms.
Compared to \mbox{\mode{GPU}}, \mbox{\mode{\proj}} achieves 34.8$\times$, 46.1$\times$, 24.4$\times$, and 31.5$\times$ on four baselines, respectively.}

\hl{Specifically, \mbox{\mode{\proj}} outperforms prior bit-sparsity accelerators, \mbox{\mode{Pragmatic}} and \mbox{\mode{BBS}}, by 4.6$\times$ and 3.2$\times$, respectively.} 
Both \mode{Pragmatic} and \mode{BBS} primarily exploit weight-only bit sparsity, which is close to 50\% in our evaluated algorithms. 
In contrast, \mode{\proj} leverages the much higher bit sparsity exposed by temporal similarity across adjacent frames, which is over 90\% in many cases (\Fig{fig:sparsity_comparison}). 
Moreover, our 1D PE-array organization along its dataflow eliminates the workload imbalance in prior bit-sparsity accelerators. 
As a result, \mode{\proj} achieves higher PE utilization than both \mode{Pragmatic} and \mode{BBS}.

\hl{We also compare \mbox{\mode{\proj}} against recent diffusion accelerators, e.g., \mbox{\mode{Cambricon-D}}, \mbox{\mode{Ditto}}, and \mbox{\mode{Exion}}. 
On average, \mbox{\mode{\proj}} achieves 6.1$\times$, 3.3$\times$, and 4.3$\times$ speedup over these designs, respectively. }
Among these baselines, \mode{Ditto} is the most closely related to our approach, as it exploits similarity across adjacent diffusion timesteps to skip redundant computation. 
As a result, \mode{Ditto} achieves higher speedup than both \mode{Cambricon-D} and \mode{Exion}. 
In contrast, \mode{Exion} focuses on sparsity in attention computation. 
However, as shown in \Fig{fig:exec_breakdown_attn}, attention contributes only a small fraction of the total execution time in diffusion-based VLA. 
Thus, \mode{Exion} delivers the lowest speedup among the three baselines.

Overall, \mode{\proj} achieves the highest speedup for two reasons. 
First, it exploits temporal-aware bit sparsity to eliminate more redundant computation than prior approaches. 
Second, its speculative inference alleviates the memory pressure that arises after computation is substantially reduced. 
Specifically, once the computation workload is significantly lowered, the execution becomes memory-bound, and speculative inference can mitigate this by amortizing overall off-chip memory access latency.

\begin{figure}
    \centering
    \includegraphics[width=\columnwidth]{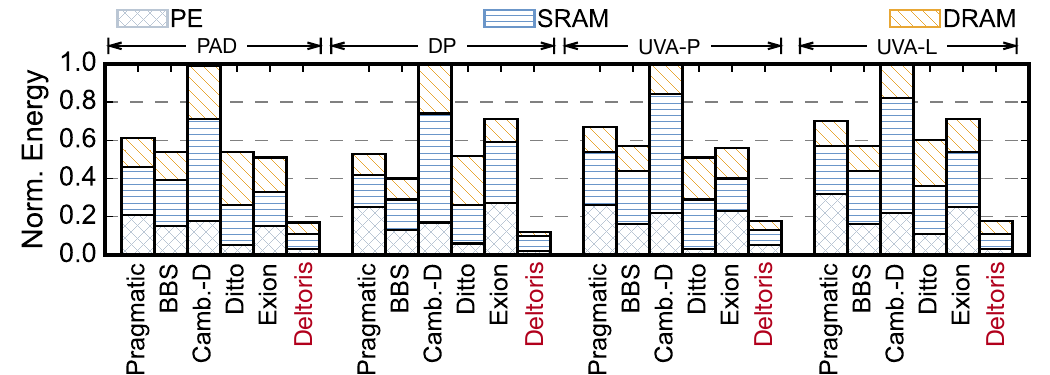}
    \caption{\hl{Normalized energy breakdown of different hardware baselines. All results are normalized to \mbox{\mode{Cambricon-D}}.}}
    \label{fig:energy_breakdown}
\end{figure}

\para{Energy Savings.}
\Fig{fig:energy} compares the energy savings of different accelerators.
Here, we define energy saving as the ratio between the energy consumption of the baseline, i.e., \mode{GPU}, and that of the corresponding hardware accelerator.
Across all three evaluated algorithms, \mode{\proj} consistently achieves the highest energy efficiency.
\hl{In particular, \mbox{\mode{\proj}} delivers up to 850$\times$, 1040$\times$, 746$\times$ and 652$\times$ energy savings on PAD, DP, UVA-P, and UVA-L respectively.}

\Fig{fig:energy_breakdown} further shows the detailed energy breakdown of different hardware baselines, where all results are normalized to \mode{Cambricon-D}. 
As shown in \Fig{fig:energy_breakdown}, the large energy savings of \mode{\proj} come from jointly reducing both computation and data traffic. 
Compared to the other two bit-serial accelerators, \mode{\proj} achieves lower DRAM energy due to two contributions: speculative inference and data compression.
Speculative inference amortizes off-chip data traffic across multiple frames, while data compression further reduces communication volume. 
At the same time, our temporal-aware bit-sparsity algorithm further lowers both SRAM-access energy and PE compute energy by eliminating redundant bit-level operations. 
A similar trend can be observed for \mode{Ditto}, which also exploits similarity across diffusion timesteps and thus significantly reduces PE compute energy. 
In contrast, \mode{Cambricon-D} must perform at least half-precision computation, i.e., 4-bit, on every input element, resulting in substantially higher compute energy.

\para{Operation Reduction.}
\Fig{fig:ops} shows the normalized operation counts of different hardware baselines.
All results are normalized to \mode{GPU}.
\hl{On average, \mbox{\mode{\proj}} achieves the highest operation reduction across all evaluated VLA models with an average of 93\% of operation reduction.}
The reduction primarily comes from our temporal-aware bit-level sparsity, which eliminates redundant computations across control steps.
In contrast, prior bit-sparsity accelerators, e.g., \mode{Pragmatic} and \mode{BBS}, only reduce around 30\% to 50\% by leveraging weight sparsity.
\mode{Ditto} also achieves relatively high operation reduction by exploiting the similarity between denoising steps, but still requires more computations compared to \mode{\proj}.
Overall, the result shows that frame-wise similarity provides the highest operation reductions compared to other techniques.

\subsection{Ablation Study}
\label{sec:eval:abl}

\begin{figure}
    \centering
    \includegraphics[width=\columnwidth]{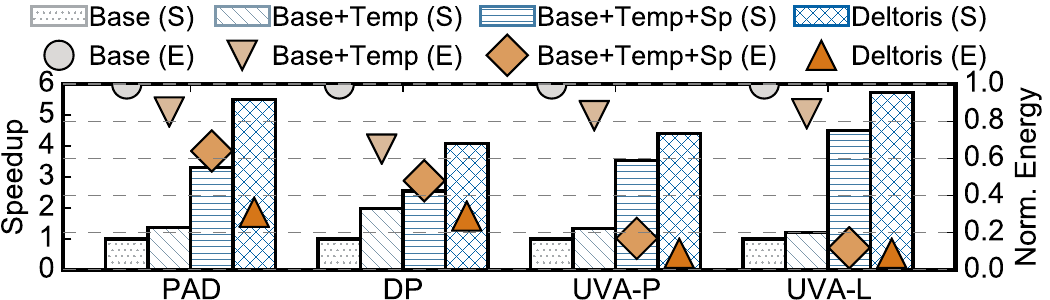}
    \caption{\hl{Ablation study of \mbox{\proj}. We compare different configurations, including \mbox{\mode{Base}}: baseline hardware with weight-only sparsity; \mbox{\mode{Base+Temp}}: Baseline hardware with temporal-aware bit sparsity; \mbox{\mode{Base+Temp+Sp}}: baseline hardware with temporal-aware bit sparsity and speculative inference; \mbox{\mode{\proj}}: the full-fledged design. (S): speedup; (E): normalized energy. UVA-P: UVA result on the PushT dataset; UVA-L: UVA result on the LIBERO dataset. Numbers are normalized to the corresponding \mbox{\mode{Base}}.}}
    \label{fig:ablation}
\end{figure}

We perform an ablation study to quantify the individual contributions of \proj.
\Fig{fig:ablation} shows both the speedup and normalized energy of four design variants.
\mode{Base} is a variant that adopts our hardware architecture but with only weight-level sparsity.
\mode{Base+Temp} then augments \mode{Base} with temporal-aware bit sparsity.
\mode{Base+Temp+Sp} further incorporates speculative inference.
Finally, \mode{\proj} is a variant that represents the full-fledged design with all proposed optimizations enabled.

\hl{The overall result shows that temporal-aware bit sparsity is the primary contributor to computation reduction.
That's why \mbox{\mode{Base+Temp}} achieves 1.5$\times$ speedup compared to \mbox{\mode{Base}}.
Speculative inference further improves performance and energy efficiency by amortizing the off-chip data traffic, since the workload of \mbox{\mode{Base+Temp}} is already memory-bound.
Thus, \mbox{\mode{Base+Temp+Sp}} achieves 3.5$\times$ speedup and 21\% energy reduction compared to \mbox{\mode{Base}}.
\mbox{\mode{\proj}} further reduces the off-chip traffic via our data compression scheme to achieve the highest overall performance and energy efficiency.}

\begin{figure}[t]
\centering
\begin{minipage}[t]{0.48\columnwidth}
  \centering
  \includegraphics[width=\columnwidth]{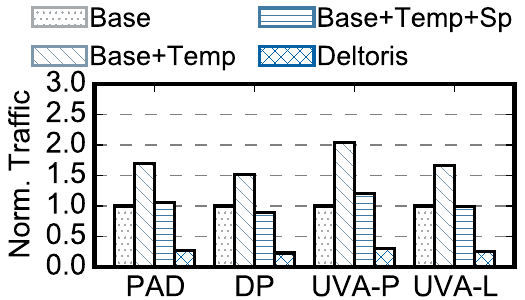}
  \caption{\hl{Ablation study on the off-chip data traffic of different hardware variants.}}
  \label{fig:ablation_traffic}
\end{minipage}
\hspace{2pt}
\begin{minipage}[t]{0.48\columnwidth}
  \centering
  \includegraphics[width=\columnwidth]{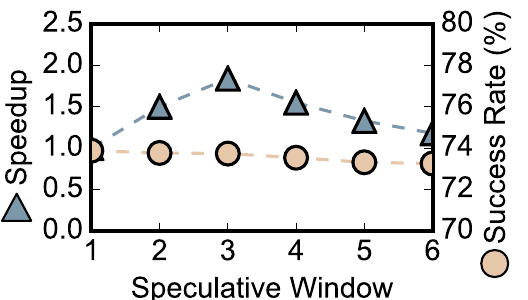}
  \caption{The sensitivity of performance and accuracy to speculative window.}
  \label{fig:speculative_sens}
\end{minipage}
\end{figure}

\Fig{fig:ablation_traffic} further analyzes the off-chip data traffic across different design variants.
Compared to \mode{Base}, \mode{Base+Temp} increases data traffic due to additional intermediate data accesses introduced by temporal-aware bit-sparsity operations.
By introducing speculative inference, \mode{Base+Temp+Sp} reduces the off-chip data traffic via amortized data loading across multiple frames.
Finally, \mode{\proj} further reduces off-chip traffic via data compression.
  
\subsection{Sensitivity Study}
\label{sec:eval:sen}

% \begin{figure}[t]
% \centering
% \begin{minipage}[t]{0.48\columnwidth}
%   \centering
%   \includegraphics[width=\columnwidth]{speedup_pe_buf}
%   \caption{The sensitivity of performance to PE array size and on-chip buffer size.}
%   \label{fig:pe_buf_sens}
% \end{minipage}
% \hspace{2pt}
% \begin{minipage}[t]{0.48\columnwidth}
%   \centering
%   \includegraphics[width=\columnwidth]{energy_pe_buf}
%   \caption{The sensitivity of performance to PE array size and on-chip buffer size.}
%   \label{fig:pe_buf_sens}
% \end{minipage}
% \end{figure}

\begin{figure}[t]
\centering
\subfloat[Speedup.]{
    \label{fig:pe_buf_speedup}
    \includegraphics[width=0.48\columnwidth]{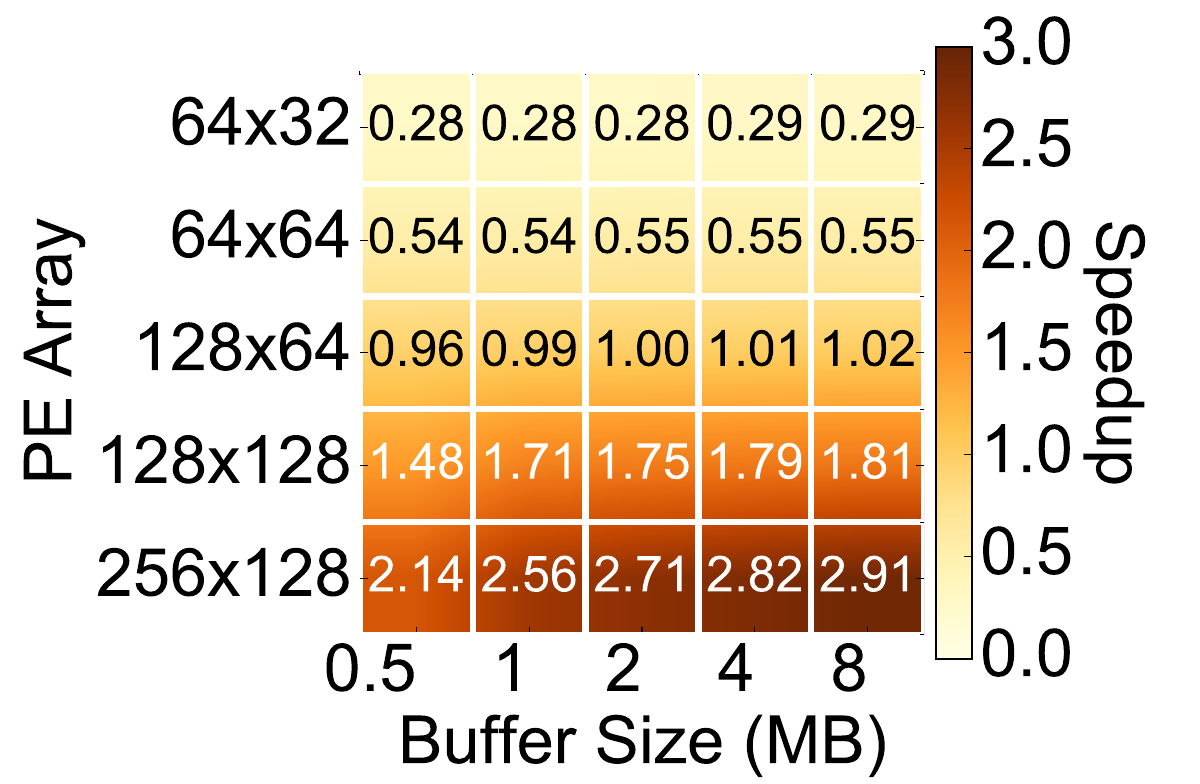}}
\subfloat[Normalized energy.]{
    \label{fig:pe_buf_energy}
    \includegraphics[width=0.48\columnwidth]{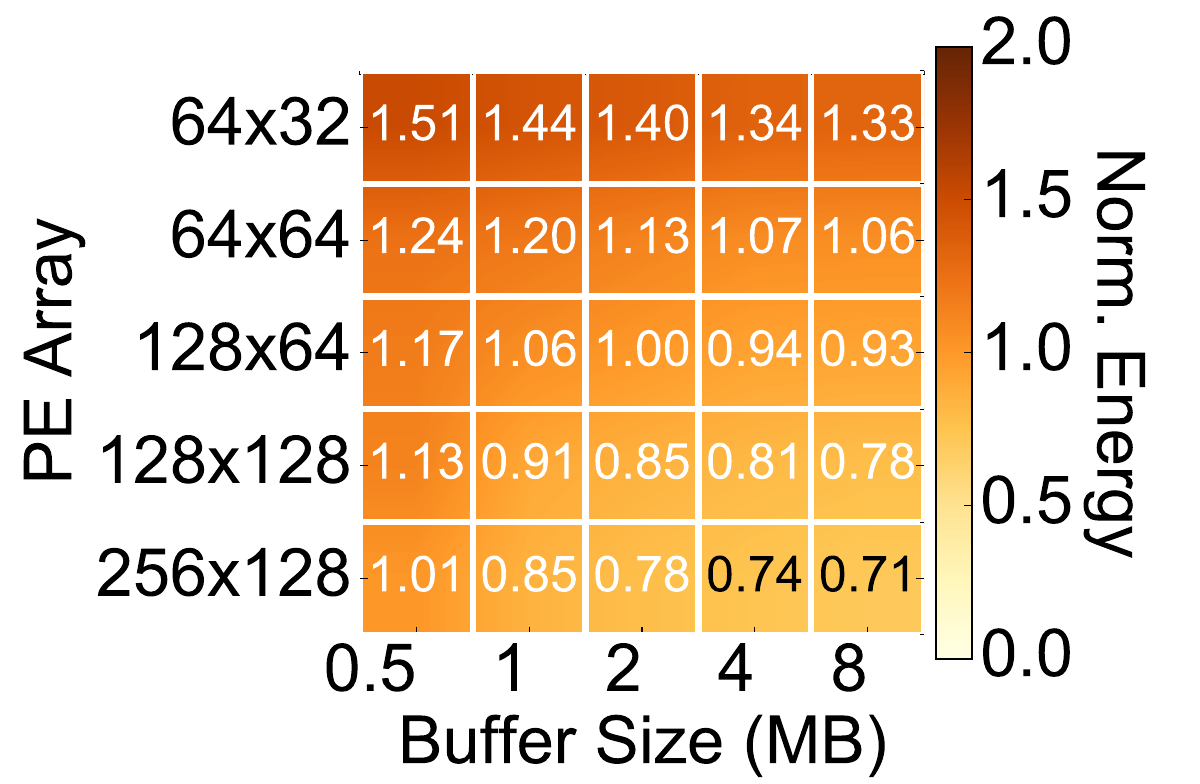}}
\caption{The sensitivity of performance and normalized energy to PE array size and on-chip buffer size.}
\label{fig:pe_buf_sens}
\end{figure}

\para{Speculative Window.}
We further study the impact of speculative window size on performance and accuracy.
Here, we use PAD as a case study.
\Fig{fig:speculative_sens} shows the speedup and success rate as we sweep the size of the speculative window.
As the window size increases, the speedup first increases, then decreases.
This is because a larger window increases the prediction overhead, while the fraction of accepted speculative candidates gradually decreases.
Meanwhile, the success rate slightly decreases as the window size increases.
% This indicates that speculative inference is robust.
Overall, a speculative window size of 3 is a good tradeoff.

\para{PE \& Buffer Size.}
\Fig{fig:pe_buf_sens} shows the sensitivity of \proj to different PE array sizes and on-chip buffer sizes.
We also use the PAD algorithm to illustrate the trend and normalized all the numbers to our default configuration, 128 PE arrays, each with 64 PEs ($128 \times 64$).
Overall, increasing the PE array size improves performance by enabling higher parallelism, thus higher performance.
On the other hand, simply increasing the on-chip buffer size will not improve the performance significantly.
However, increasing buffer sizes reduces off-chip traffic and improves energy efficiency.

\begin{figure}[t]
\centering
\begin{minipage}[t]{0.48\columnwidth}
  \centering
  \includegraphics[width=\columnwidth]{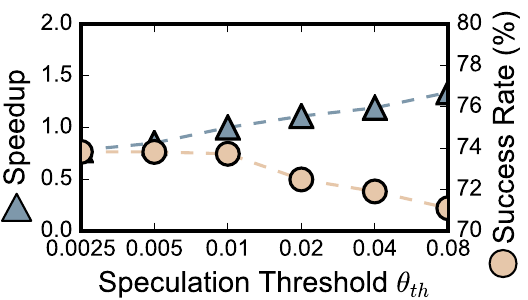}
  \caption{\hl{The sensitivity of performance and accuracy to speculation threshold, $\theta_{th}$.}}
  \label{fig:spec_threshold}
\end{minipage}
\hspace{2pt}
\begin{minipage}[t]{0.48\columnwidth}
  \centering
  \includegraphics[width=\columnwidth]{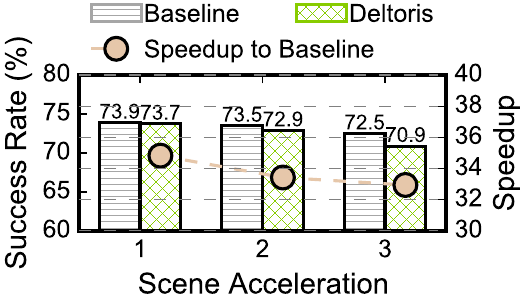}
  \caption{\hl{The accuracy of \mbox{\proj} under the scenarios with rapid motions.}}
  \label{fig:fast_motion}
\end{minipage}
\end{figure}

\para{Speculation Threshold, $\theta_{th}$.}
\hl{
\mbox{\Fig{fig:spec_threshold}} shows the sensitivity of speedup and accuracy to the speculation threshold, $\theta_{th}$, described in \mbox{\Sect{sec:spec}}.
Here, we report the results for PAD.
As shown in \mbox{\Fig{fig:spec_threshold}}, the overall speedup increases as $\theta_{th}$ becomes larger, since more speculative candidates are accepted.
Meanwhile, the success rate remains relatively stable at smaller thresholds, but starts to drop when $\theta_{th}$ reaches 0.01.
Therefore, we set $\theta_{th}$ to 0.01 in our system to balance performance and accuracy.
}

\subsection{Scenarios with Fast Motion}
\label{sec:eval:fast}

\hl{
To stress-test our technique, we further evaluate \mbox{\proj} under rapid-motion scenarios.
Since existing robotic VLA datasets do not specifically target rapid robot motions, we construct a fast-motion dataset by skipping intermediate observations during normal robot execution while allowing the robot to execute additional steps.
This increases the change between consecutive observations and effectively simulates faster robot motion.

In \mbox{\Fig{fig:fast_motion}}, we evaluate both \mbox{\proj} and the baseline algorithm on the PAD model and measure their performance and task accuracy.
The performance numbers are normalized to the GPU baseline.
The results show that, when the scene is accelerated by 2$\times$, \mbox{\proj} maintains accuracy comparable to the baseline, with less than 1\% accuracy loss.
Further increasing the motion speed may degrade the accuracy of \mbox{\proj}, as the temporal similarity between consecutive observations becomes weaker.
Meanwhile, accelerating the simulation scene would lead to modest performance degradation.
The speedup decreases from 34.8$\times$ to 33.4$\times$ when the scene is accelerated by 2$\times$.
Nevertheless, the results show that \mbox{\proj} remains robust under fast-motion scenarios.
}
\section{Related Work}
\label{sec:related}

\para{Acceleration for Robotics.}
Over the past decade, computer architecture researchers have proposed a wide range of accelerators and systems for robotics workloads~\cite{sacks2018robox, hadidi2021quantifying, liu2021micro, neuman2021robomorphic, krishnan2023autopilot, bakhshalipour2022racod, yang2023dadu, shah2023energy, neuman2023roboshape, nikiforov2023rose, shah2024collison, hao2024orianna, bakhshalipour2024tartan, choe2025ana, huang2025corki, ting2025hiper, zhang2025idea, chen2025octocache, wan2025reca, huang2026splatonic}.
Early efforts largely focused on accelerating individual robotics kernels, such as model predictive control (MPC)~\cite{sacks2018robox, bakhshalipour2022racod}, rigid-body dynamics~\cite{yang2023dadu}, localization~\cite{liu2021micro}, and collision detection~\cite{shah2023energy, shah2024collison}, each of which exposes distinct domain-specific optimization opportunities. 
A second class of work, including RoS\'E~\cite{nikiforov2023rose}, Orianna~\cite{hao2024orianna}, and AutoPilot~\cite{krishnan2023autopilot}, emphasizes system-level optimization by exploiting common bottlenecks from multiple algorithms, rather than kernel-specific acceleration. Tartan~\cite{bakhshalipour2024tartan}, which is a CPU-centric robotic processor, also falls into this category. 
More recently, architectural research has begun shifting toward embodied AI~\cite{huang2025corki, wan2025reca} and autonomous driving~\cite{chen2025octocache, choe2025ana}, where performance is governed not by a single computational kernel, but by the end-to-end interaction among perception and control.

Unlike prior robotics accelerators that focus on classical controls, \proj is the first focusing on \emph{diffusion-based VLA inference}, an emerging embodied-AI workload with new computational characteristics.
\proj is designed to exploit the strong temporal similarity across consecutive control steps in order to reduce both computation and data communication overhead during inference.

\para{Bit Sparsity.}
A large body of prior work~\cite{judd2016stripes, albericio2017bit, sharma2018bit, sharify2019laconic, lu2021distilling, lascorz2019bit, shi2024bitwave, chen2024bbs, wei2025phi} has proposed bit-serial accelerators to eliminate ineffective bit operations in DNN inference. 
Stripes~\cite{judd2016stripes} first showed that bit-serial execution can scale performance with operand precision, while Bit Fusion~\cite{sharma2018bit} proposed a fine-grained mixed-precision architecture. 
Pragmatic~\cite{albericio2017bit} and Laconic~\cite{sharify2019laconic} further improved efficiency by processing only effective terms.
Subsequent works such as Bitlet~\cite{lu2021distilling} and BitWave~\cite{shi2024bitwave} introduced various mechanisms to improve PE utilization. 
More recently, BBS~\cite{chen2024bbs} explored bidirectional bit sparsity and approximation to further reduce computation.

However, these designs primarily target weight sparsity or static bit-level inefficiency and do not exploit the temporal redundancy inherent in diffusion-based VLA workloads. 
In contrast, \proj is specifically designed to leverage temporal similarity across consecutive control steps and exposes much higher bit-level sparsity.

\para{Vision Acceleration.}
Vision workloads have long been a key topic in architecture research.
For example, both AdapTiV~\cite{yoo2024adaptiv} and HeatViT~\cite{dong2023heatvit} reduce the computational overhead of ViTs through adaptive token pruning.
ViTALiTy~\cite{dass2023vitality} and ViTCoD~\cite{you2023vitcod}, in contrast, target another key bottleneck in ViTs, the attention computation, via sparse approximation.
Recent work has also started to explore diffusion-model acceleration.
Cambricon-D~\cite{kong2024cambricon} proposes techniques to mitigate the memory overhead of differential execution, while Ditto~\cite{kim2025ditto} exploits the high similarity between adjacent denoising steps for differential computation.
EXION~\cite{heo2025exion}, on the other hand, accelerates diffusion models through sparse attention.
Meanwhile, another line of research exploits spatial and temporal correlations across video frames to avoid redundant computation~\cite{buckler2018eva2, zhu2018euphrates, ying2022exploiting, feng2019asv, zhao2020deja, zhao2021holoar, feng2024cicero, feng2025lumina, zhu2026nebula}.
These works show that temporal redundancy can be an effective way to improve efficiency.

\para{Temporal Correlation.}
\hl{
Temporal correlation has been widely studies in various vision workloads~\cite{buckler2018eva2, zhu2018euphrates, feng2019asv, ying2022exploiting, song2020vr, zhao2020deja, zhao2021holoar, xu2025vla, feng2023fast}.
Many prior studies operate directly at the pixel level and reduce computation via motion estimation~\mbox{\cite{zhu2018euphrates, feng2019asv}}.
VLA-cache~\mbox{\cite{xu2025vla}} leverages the cross-step redundancies to approximate the VLA inference.
Others~\mbox{\cite{song2020vr, song2024cmc}} exploit video codecs to guide efficient video understanding.
In addition, Deja View~\mbox{\cite{zhao2020deja}} and Cicero~\mbox{\cite{feng2024cicero}} show that spatio-temporal correlations can also improve VR efficiency.

However, these prior works primarily exploit temporal correlation through approximate computation, often at the pixel or block level.
In contrast, \mbox{\proj} exploits temporal similarity not only in the pixel space but also in the latent activation space of VLA models.
More importantly, \mbox{\proj} preserves bit-accurate inference for supported linear operators by combining temporal differential computation with bit-serial execution.
}

\para{Speculative Decoding.}
\hl{
Recent works~\mbox{\cite{wang2025spec, zheng2026kerv, zheng2026heisd, huo2025spec}} have explored speculative decoding to reduce the inference latency of AR-based VLA models.
Some studies, such as Spec-LLaVA~\mbox{\cite{huo2025spec}}, investigate different strategies for speculating future tokens.
Others, e.g., Spec-VLA~\mbox{\cite{wang2025spec}} and KERV~\mbox{\cite{zheng2026kerv}}, further incorporate robotic dynamics into the speculation process.
Although these methods demonstrate the effectiveness of speculative decoding for improving inference efficiency, they primarily target decoding-level acceleration in autoregressive models and therefore cannot be directly applied to diffusion-based VLA models.
In contrast, \mbox{\proj} is tailored for diffusion-based VLAs.
It leverages the property that diffusion-based VLAs also predict future observations together with actions, allowing a lightweight model to generate speculative future candidates that are subsequently verified by a large VLA model.
}

\section{Conclusion}
\label{sec:conc}

This paper presents \proj, a hardware-software co-designed accelerator for real-time diffusion-based VLA inference. 
We introduce two key techniques: temporal-aware bit sparsity and speculative inference to exploit the strong temporal similarity. 
To support these techniques, we design a bit-serial architecture with a novel PE-array. 
Together, these contributions take the first step toward enabling real-time VLA processing for embodied AI systems.

\begin{acks}

This work was supported by the Fundamental and Interdisciplinary Disciplines Breakthrough Plan of the Ministry of Education of China (JYB2025XDXM113), the National Natural Science Foundation of China (NSFC) Grants (62532006 and 62402312), Shanghai Pujiang Talent Program (24PJA044), and Shanghai Qi Zhi Institute Innovation Program (SQZ202316).

\end{acks}

% \clearpage

%%%%%%% -- PAPER CONTENT ENDS -- %%%%%%%%

\bibliographystyle{ACM-Reference-Format}
\bibliography{refs}

\end{document}